\documentclass[aps,prd,twocolumn,
eqsecnum,showpacs
]{revtex4}

\usepackage{amsmath,amssymb,amsfonts,
color,graphicx,latexsym,theorem,mathrsfs
}

\newcommand{\ma}[1]{\mbox{$\mathcal{#1}$}}
\newcommand{\mas}[1]{\mbox{$\mathscr{#1}$}}
\newcommand{\D}{{\rm d}}
\newcommand{\ti}{\tilde}
\newcommand{\we}{\wedge}

\begin{document}

\title{
Cosmological rotating black holes in five-dimensional 
fake supergravity 
}

\author{Masato {\sc Nozawa}$^1$}
\email{nozawa@gravity.phys.waseda.ac.jp}
\author{Kei-ichi {\sc Maeda}$^{1,2}$}
\email{maeda@waseda.jp}


\address{$^1$ Department of Physics, Waseda University, 
Okubo 3-4-1, Shinjuku, Tokyo 169-8555, Japan\\
$^2$ Waseda Research Institute for Science and Engineering,
Okubo 3-4-1, Shinjuku, Tokyo 169-8555, Japan}

\date{\today}

\begin{abstract} 
In recent series of papers, we found an arbitrary dimensional, 
time-evolving and spatially-inhomogeneous solutions in  
Einstein-Maxwell-dilaton gravity with particular couplings. 
Similar to the supersymmetric case the solution can be arbitrarily 
superposed in spite of non-trivial time-dependence, since
the metric is specified by a set of harmonic functions. 
When each harmonic has a single point source at the center, 
the solution describes a spherically symmetric black hole with regular
Killing horizons and the spacetime 
approaches asymptotically to the Friedmann-Lema\^itre-Robertson-Walker (FLRW) cosmology. 
We discuss in this paper that in 5-dimensions this equilibrium condition traces back
to the 1st-order ``Killing spinor'' equation in ``fake supergravity'' 
coupled to arbitrary ${\rm U}(1)$ gauge fields and scalars. 
We present a 5-dimensional, asymptotically FLRW,  rotating black-hole
solution admitting a nontrivial ``Killing spinor,''
which is a spinning generalization of our previous solution.    
We argue that the solution admits nondegenerate and rotating 
Killing horizons in contrast with the supersymmetric solutions. It is shown that 
the present pseudo-supersymmetric solution admits closed timelike
curves around the central singularities.  
When only one harmonic is time-dependent, 
the solution oxidizes to 11-dimensions and realizes the dynamically 
intersecting M2/M2/M2-branes in a rotating Kasner universe. The Kaluza-Klein
type black holes are also discussed.     
\end{abstract}


\maketitle


\section{Introduction}

Supersymmetric solutions in supergravity 
have played an important r\^ole in the development of string theory 
and the anti-de Sitter(AdS)/conformal field theory
(CFT)-correspondence. A pioneer work in this direction was the 
great success of microscopic deviation of black hole entropy from
the viewpoint of intersecting D-branes.  
By virtue of the saturation of Bogomol'nyi-Prasad-Sommerfield (BPS) bound, 
the supersymmetric solutions can provide arena for exploring 
the non-perturbative limits of string theory. 
The BPS equality constraints the supersymmetry variation spinor  
to satisfy the 1st-order differential equation. 
Such a covariantly constant spinor is called a Killing spinor, which ensures that the 
energy is positively bounded by central charges, guaranteeing the
stability of the theory.  The relationship between vacuum stability and
BPS states was suggested by Witten's positive energy theorem~\cite{Witten:1981mf}, and
later validated firmly by~\cite{GH1982,PET}.

From a standpoint of pure gravitating object, black hole solutions 
admitting a Killing spinor are sharply distinguished from non-BPS black
hole solutions. These BPS configurations are dynamically very simple.   
First of all, BPS black hole solutions necessarily have zero Hawking
temperature (the converse is not true), implying  that 
the horizon is degenerate.  Accordingly they are free
from thermal excitation. Such a non-bifurcating horizon 
universally admits a throat infinity and  
enhanced isometries of ${\rm SO}(2, 1)$~\cite{Kunduri:2007vf}. 
Secondly, most BPS solutions satisfy ``no-force'' condition. For example, we
are able to superpose the extreme Reissner-Nordstr\"om solutions at our
disposal due to the delicate compensation between the gravitational attractive
force and the electromagnetic repulsive force. The resulting multicenter metric, 
originally found by Majumdar and Papapetrou, maintains static
equilibrium and describes collection of charged black holes~\cite{Hartle:1972ya}. 
This property can be ascribed  to the complete linearization of field
equations. Besides these, all the BPS black holes are known to be 
strictly stationary, viz, the ergoregion does not exist even if the 
black hole has nonvanishing angular momentum. Dynamically evolving
states are not compatible with supersymmetry.

Then, to what extent these known intuitive properties continue to hold?
Motivated by this inquiry 
it is important to explore general properties and classify BPS solutions.  
A first progress was made by 
Tod,  who catalogued all the BPS solutions admitting nontrivial
Killing spinors of 4-dimensional $\ma N=2$ supergravity~\cite{Tod}, inspired by the
early study of Gibbons and Hull~\cite{GH1982}.
Recently, Gauntlett {\it et al.}~\cite{GGHPR} 
were able to obtain general supersymmetric solutions in 5-dimensional
minimal supergravity exploiting bilinears constructed from a Killing
spinor. Since their technique has no restriction upon the spacetime
dimensionality,  reference~\cite{GGHPR} has sparked  a considerable development in the 
classifications of supersymmetric solutions in various  
supergravities~\cite{Gauntlett:2003fk,Gutowski:2003rg,Caldarelli:2003pb,GR}.
This formalism is useful for finding supersymmetric black holes~\cite{Gutowski:2004ez,Kunduri:2006ek}
and black rings~\cite{Elvang:2004rt,Elvang:2004ds,Gauntlett:2004qy}, and
for proving uniqueness theorem of certain black holes~\cite{Reall2002,Gutowski:2004bj}.
It turns out that all the BPS black holes fulfill above mentioned
properties except for the equilibrium condition which is valid
only in the ungauged case.

On the other hand,  non-BPS black hole solutions--especially the 
time-dependent black hole solutions--have been much less understood. 
In this paper we address some properties of cosmological black-hole
solutions which have an interpretation as arising from the gauged
supergravity with non-compact R-symmetry gauged. The most simplest
theory is the 4-dimensional minimal de Sitter supergravity consisting of  
the graviton, the Maxwell fields and a {\it positive} cosmological constant~\cite{Pilch:1984aw}.  
A time-dependent solution in this theory was found by Kastor and Traschen~\cite{KT,London},
which is the generalization of Majumdar-Papapetrou solution in the 
de Sitter background. The Kastor-Traschen solution describes 
coalescing black holes in the contracting de Sitter universe (or
splitting white holes in the expanding de Sitter universe) and
inherits some salient characteristics from the Majumdar-Papapetrou solution. 
The reason why multicenter metric is in mechanical equilibrium 
irrespective of the time-dependence is attributed to the
first-order ``BPS equation'' that extremizes the action, 
allowing the complete linearization of field equations. Since 
these ``BPS'' states are not truly supersymmetric in the usual sense,   
they are referred to as pseudo-supersymmetric and the corresponding
theory is called a ``fake'' supergravity. 
Recently, all pseudo-supersymmetric solutions in 4- and
5-dimensional fake de Sitter fake supergravity were classified using the
spinorial geometry method~\cite{Gutowski:2009vb,Grover:2008jr}
(see~\cite{Meessen:2009ma} for a non-Abelian generalization).

In this paper, we discuss properties of pseudo-supersymmetric solutions
of 5-dimensional fake supergravity with arbitrary number of 
${\rm U}(1)$ gauge fields and scalar fields. Some time-dependent black
hole solutions in this theory have been available so 
far~\cite{Behrndt:2003cx,Klemm:2000gh}, but their properties and
causal structures are yet to be explored.  Even for the simplest case in
which the harmonic function is sourced by a single point mass, 
the spacetime is highly dynamical except in the de Sitter supergravity. 
In the present case the background spacetime is the 
Friedmann-Lema\^itre-Robertson-Walker (FLRW) cosmology. 
(In the context of fake supergravity, 
it is argued that the FLRW cosmologies are duals of supersymmetric
domain walls. See~\cite{DW_FRW} for details.)  
A series of recent papers of present authors~\cite{MN,MNII} revealed 
that the solution of a single point source  
found in~\cite{MOU,GMII} actually describes a charged black hole in the  
FLRW cosmology.   Though the metric in~\cite{MOU,GMII} were shown 
to be the  exact solutions of Einstein-Maxwell-dilaton system,  
we show in this paper that the 5-dimensional solutions
of~\cite{MOU,MNII} in fact satisfy the 1st-order BPS equation in fake
supergravity. The pseudo-supersymmetry is indeed consistent with an
expanding universe.  
This work will establish new insights for black holes in 
time-dependent and non-supersymmetric backgrounds.

The main concern in this paper is to see the effects of black-hole rotation 
in 5-dimensions by restricting to the single point mass case. 
As it turns out, rotation makes the properties of spacetime much richer. 
Our work is organized as follows. In the next section we describe
a fake supergravity model and derive (in a gauge different
from~\cite{Klemm:2000vn}) a rotating, time-dependent solution
preserving the pseudo-supersymmetry.   Section~\ref{sec:bhs} is devoted
to explore physical and geometrical properties of the spacetime. 
We establish that the black hole horizon is generated by a rotating 
Killing horizon, in sharp contrast with the supersymmetric black-hole
horizon which admits a non-rotating degenerate Killing horizon without
an ergoregion.   
It is also demonstrated that the solution generally admits closed timelike
curves in the vicinity of timelike singularities (with trivial
fundamental group).  
Combining the analysis of the near-horizon geometries, we 
shall elucidate the causal structures by illustrating Carter-Penrose diagrams. 
In section~\ref{sec:liftup} the liftup and reduction scheme of the
5-dimensional solution is accounted for. It is shown that the
5-dimensional solutions derived in \cite{MOU,MNII} and in  section~\ref{sec:bhs}
are elevated to describe the non-BPS dynamically
intersecting M2/M2/M2-branes in 11-dimensional supergravity. 
Upon dimensional reduction the 4-dimensional black hole~\cite{MOU,GM} is
obtainable. We shall also present some Kaluza-Klein black holes in the FLRW
universe.   Section~\ref{conclusion} gives final remarks.

We will work in mostly plus metric signature and the standard 
curvature conventions 
$2\nabla_{[\rho }\nabla_{\sigma] }V^\mu={R^\mu}_{\nu\rho\sigma}V^\nu$. 
Gamma matrix conventions are such that   
$\gamma_{\mu\nu\rho\sigma\tau }=i\epsilon_{\mu\nu\rho\sigma\tau }$
with $\epsilon_{01234}=1$ and $\bar \psi:=i\psi^\dagger \gamma^0$.


\section{Five dimensional solutions in minimal supergravity}
\label{5D_sol_SG}

The metrics obtained in~\cite{MOU,GMII,MN,MNII} are the exact solutions
of Einstein's equations sourced by two-${\rm U}(1)$ fields and a scalar
field coupled to the gauge fields.  Since the solution involves two kinds of harmonic
functions, it manifests mechanical equilibrium regardless of
time-evolving spacetime. When each harmonic has a point source at the center, 
the solution in~\cite{MOU,GMII,MN,MNII} describes a spherically
symmetric black hole embedded in the FLRW cosmology. 
In this section, we consider 
a five-dimensional supergravity-type Lagrangian and present more
general (pseudo) BPS solutions, which encompass the
5-dimensional solution in~\cite{MOU,MNII} as special limiting cases.

Let us start from the 
minimal 5-dimensional gauged supergravity coupled to 
$N$ abelian vector multiplets. The bosonic 
action involves graviton, ${\rm U}(1)$ gauge fields $A^{(I)}$
($I=1,..., N$) 
with real scalars $\phi^A$ ($A=1, ..., N-1$)~\cite{Gunaydin:1984ak},
\begin{align}
S
=&\frac{1}{2\kappa_5^2 }\int \left[\left({}^5 R 
+2 \mathfrak g^2 V \right)\star _51
-\ma G_{AB} \D \phi^A \we \star_5 \D \phi^B
\right. \nonumber \\ & \left.
- G_{IJ} F^{(I)}\we \star_5F^{(J)}-\frac 16 C_{IJK}
A^{(I)}\we F^{(J)}\we F^{(K)}
\right]\,,
\label{5Daction}
\end{align}
where $F^{(I)}=\D A^{(I)}$ are the field strengths of gauge fields and
$\mathfrak g$ is the coupling constants corresponding to the 
reciprocal of the AdS curvature radius.  
$C_{IJK}$ are 
constants symmetric in $(IJK)$ and obey the ``adjoint identity''
\begin{align}
 C_{IJK}C_{J'(LM}C_{PQ)K'}\delta ^{JJ'}\delta ^{KK'}
=\frac 43 \delta _{I(L}C_{MPQ)}\,,
\end{align}
where the round brackets denote symmetrization of the suffixes.
The potential $V$ can be expressed in terms of a superpotential $W$ as 
\begin{align}
V=6W^2-\frac 92 \ma G^{AB} (\partial_A W)( \partial_B W)\,,
\end{align}
where $\partial_AX^I:=\D X^I(\phi)/\D \phi^A$. 
The superpotential takes the form,
\begin{align}
 W=V_IX^I\,,
\end{align}
where $V_I$ are constants arising from an Abelian gaugings of the 
${\rm SU}(2)$-R symmetry with the gauge field $A=V_IA^I$~\cite{Gunaydin:1984ak}. 
The $N$-scalars $X^I$ are constrained by
\begin{align}
\ma V:= \frac 16C_{IJK}X^IX^JX^K =1\,. 
\label{constraint_XI}
\end{align} 
It is convenient to define 
\begin{align}
 X_I= \frac 16 C_{IJK}X^JX^K\,,
\end{align}
in terms of which Eq.~(\ref{constraint_XI})
is simply $X_IX^I=1$. 
The coupling matrix 
$G_{IJ}$ is the metric of the 
``very special geometry''~\cite{deWit:1992cr} defined by
\begin{align}
 G_{IJ}:=-\frac 12\left.\frac{\partial^2  }{\partial X^I\partial X^J}\ln \ma V
\right|_{\mathcal V=1}=
\frac 92 X_IX_J-\frac 12 C_{IJK}X^K\,,
\label{G_IJ}
\end{align} 
with its inverse
\begin{align}
 G^{IJ}=2X^IX^J-6 C^{IJK}X_K\,,
\end{align}
where 
$C^{IJK}=\delta^{IL}\delta ^{JP}\delta^{KQ}C_{LPQ}$. 
The other coupling matrix $\ma G_{AB}$ is given by
\begin{align}
 \ma G_{AB}=G_{IJ} \partial _AX^I\partial_B X^J\,,
\end{align}
It follows that 
\begin{align}
 X^I= \frac 92C^{IJK}X_JX_K\,, 
\end{align}
and 
\begin{align}
 X_I=\frac 23 G_{IJ}X^J\,, \qquad
 X^I=\frac 32 G^{IJ}X_J\,.
\end{align}
From these relations, we obtain useful  expressions
\begin{align}
 &\D X_I=-\frac 23G_{IJ}\D X^J\,,\qquad
\D X^I=-\frac 32 G^{IJ}\D X_J\,, \nonumber \\
&
X^I\D X_I=X_I\D X^I=0\,, \nonumber \\
&\ma G^{AB}\partial_A X^I\partial_BX^J=G^{IJ}-\frac 23X^IX^J.
\label{useful_relation}
\end{align}
Using these formulae, the potential reads
\begin{align}
 V=27C^{IJK}V_IV_JX_K\,.
\end{align}
If this theory is derived via gauging the supergravity derived from
the Calabi-Yau compactification of M-Theory, $\ma V$ is 
the intersection form, $X^I$ and $X_I$ correspond respectively to
the size of the two- and four-cycles. The constants $C_{IJK}$ are the 
intersection numbers of the Calabi-Yau threefold and  $N$ denotes the 
Hodge number $h_{1,1}$~\cite{Papadopoulos:1995da}.

\begin{widetext}
The governing equations are the Einstein equations (varying $g^{\mu\nu }$),
\begin{align}
{}^5R_{\mu \nu } -\frac 12 \left({}^5 R +2\mathfrak g^2 V\right)
 g_{\mu \nu } =G_{IJ}
\left[(\nabla_\mu X^I)(\nabla_\nu X^J)-\frac 12 
(\nabla^\rho X^I)(\nabla_\rho X^J)g_{\mu\nu }
+F^{(I)\rho }_\mu F^{(J)}_{\nu\rho }-\frac 14
F^{(I)}_{\rho\sigma }F^{(J)\rho\sigma }g_{\mu \nu }
\right]\,,
\end{align}
the electromagnetic field equations (varying $A^{(I)}$),
\begin{align}
\nabla_\nu \left(
G_{IJ}F^{(J)\mu \nu }
\right)-\frac{1}{16}C_{IJK}\epsilon^{\mu\nu\rho\sigma\tau  }
F^{(J)}_{\nu\rho }F^{(K)}_{\sigma\tau }=0\,,
\label{Maxwell_eq}
\end{align}
where $\epsilon_{\mu\nu\rho\sigma\tau }$
is the metric-compatible volume element, 
and the scalar field equations (varying $\phi^A$),
\begin{align}
&\biggl[
\nabla^\mu \nabla_\mu X_I+
6\mathfrak g^2 V_LV_MC_{IJK}C^{KLM}X^J\nonumber \\
&~~~
+\left(C_{IJL}X_KX^L-\frac 16C_{IJK}\right)
\left((\nabla_\mu X^J)(\nabla^\mu X^K)
+\frac 12 F^{(J)}_{\mu\nu }F^{(K)\mu\nu }\right)
\biggr]
\partial_A X^I=0\,.
\label{scalar_eq}
\end{align}
From the condition $X_I\D X^I=0$, the terms in square-bracket in the
above equation must be proportional to $X_I$. Denoting it by $LX_I$,  
one obtains the expression of $L$ using the relation~$X_IX^I=1$. 
Then  the scalar equations are rewritten as  
\begin{align}
\nabla^\mu \nabla_\mu X_I+
\left(\frac 12C_{JKL}X_IX^L-\frac 16C_{IJK}\right)
(\nabla^\mu X^J) (\nabla_\mu X^K)
+6\mathfrak g^2C^{JLM}V_LV_M
\left(6X_IX_J-C_{IJK}X^K\right)&\nonumber \\
+\frac 12 
\left(C_{IJL}X_KX^L-\frac 16 C_{IJK}-6X_IX_JX_K+\frac
 16C_{JKL}X_IX^L\right)F^{(J)}_{\mu\nu }
F^{(K)\mu\nu }
&=0\,.
\end{align}

The supersymmetric transformations for the gravitino $\psi_\mu $ and 
gauginos $\lambda_A $ are given by 
\begin{align}
\delta \psi_\mu 
&=\left[\ma D_\mu -\frac {3i}2 \mathfrak g V_IA^{(I)}_\mu
+\frac{i}{8}X_I\left({\gamma _\mu }^{\nu\rho }
-4{\delta_\mu }^\nu \gamma^\rho \right)F_{\nu\rho }^{(I)}
+\frac 12 \mathfrak g \gamma_\mu X^IV_I\right]\epsilon \,, 
\label{KS_STU1}\\
\delta \lambda _A &= \left[\frac{3}{8}\gamma^{\mu\nu }F_{\mu\nu }^{(I)}
\partial_A X_I
-\frac i{2}\ma G_{AB}\gamma ^\mu \partial_\mu \phi^B 
+\frac{3i}2 \mathfrak g V_I\partial_A X^I 
\right]\epsilon \,,
\label{KS_STU2}
\end{align}
\end{widetext}
where $\epsilon $ is a spinor generating an infinitesimal supersymmetry
transformation. 
Here and throughout the paper, 
$\ma D_\mu $ will be used for a gravitationally-covariant derivative
defined by 
\begin{align}
 \ma D_\mu \epsilon =\left(\partial_\mu +\frac 14{\omega _\mu
 }^{ab}\gamma_{ab} \right)\epsilon \,,
\end{align}
where ${\omega_\mu }^{ab}$ is a spin-connection without torsion.
We have used the Dirac spinor
instead of the symplectic-Majorana spinor.  
The supersymmetric solutions in this theory have been
analyzed~\cite{GR}. One recovers ungauged supergravity
by $\mathfrak g\to 0$.

\subsection{Pseudo-supersymmetric solutions in fake supergravity}

If we consider a non-compact gauging of R-symmetry,
an imaginary coupling arises, $\mathfrak g \to i k ~(k\in \mathbb R)$. 
Since only the R-symmetry is gauged, the imaginary coupling reflects the
non-compactness of R-symmetry. The Lagrangian~(\ref{5Daction}) is neutral 
under the R-symmetry, so that the theory is free from
the ghost-like contribution.  
This theory is called a fake supergravity.
The fake ``Killing spinor'' equations reduce to
\begin{widetext}
\begin{align}
\left[\ma D_\mu+\frac {3k}2  V_IA^{(I)}_\mu +\frac{i}{8}X_I\left({\gamma _\mu }^{\nu\rho }
-4{\delta_\mu }^\nu \gamma^\rho \right)F_{\nu\rho }^{(I)}
+\frac i2 k  \gamma_\mu X^IV_I\right]\epsilon =0
\label{fakeKS1}\,, \\
\left[\frac{3}{8}\gamma^{\mu\nu }F_{\mu\nu }^{(I)}
\partial_A X_I
-\frac i{2}\ma G_{AB}\gamma ^\mu \partial_\mu \phi^B 
-\frac{3}2 k V_I\partial_A X^I 
\right]\epsilon =0 \,.\label{fakeKS2}
\end{align}
\end{widetext}
Here, the supercovariant derivative operator is no longer
hermitian for $k\in \mathbb R$. This implies that we are unable to use 
$\epsilon $ to prove the positive energy theorem in the usual manner. Still, we presume
that the above equations~(\ref{fakeKS1}) and (\ref{fakeKS2}) continue to
be valid for $k\in \mathbb R$.

Inferring from the supersymmetric solutions in~\cite{GR}, 
we assume the standard metric ansatz, 
\begin{align}
 \D s_5^2 =-f^2 (\D t+\omega )^2 +f^{-1}h_{mn}\D x^m \D x^n\,,
\label{SUSYmetric}
\end{align}
where the 4-metric $h_{mn}$  is orthogonal to 
$V^\mu =(\partial/\partial t)^\mu $ ($i_Vh_{\mu\nu }=0$)
and supposed to be independent of $t$ ($\mas L_V h_{\mu\nu}=0$). 
The one-form  $\omega $ corresponds to the ${\rm U}(1)$
fibration of the transverse base space $(\ma B, h_{mn})$.  
In what follows, indices $m, n,...$ are raised and lowered by $h_{mn}$ and its inverse
$h^{mn}$. The connection $\omega $ is orthogonal to the timelike vector
field $V^\mu $ and assumed to be independent of $t$ 
($\mas L_{V}\omega =0$).
We further suppose that the lapse function is given by  
\begin{align}
f^{-3}=\frac 16C^{IJK}H_IH_JH_K\,,
\label{f_cube}
\end{align}
where $H_I$'s are some functions. 
We also assume the profiles of the   
electromagnetic and the scalar fields as  
\begin{align}
 A^{(I)} =fX^I (\D t+\omega )\,, \qquad 
 X_I =\frac 13 f H_I\,. 
\label{SUSYgauge}
\end{align}
In the ungauged supersymmetric case (when $\mathfrak g=0$),  
the condition~(\ref{f_cube}) is obtained as a special case of
 the general supersymmetric
solutions, as referred to hereinafter in section~\ref{sec:liftup}. 
In this section, we just assume (\ref{f_cube}).

Taking the orthonormal frame 
\begin{align}
e^0=f(\D t+\omega ) \,, \qquad e^i =f^{-1/2} \hat e^i\,, 
\end{align}
where $\hat e^i$ is the orthonormal frame for $h_{mn}$, one can calculate
the time and spatial components of ``Killing spinor'' equation (\ref{fakeKS1}), which are
given by
\begin{widetext}
\begin{align}
& \biggl[\partial_t
 +kfV_IX^I+\left\{\frac{1}{2}fkV_IX^I+\frac{1}{4}f^3\partial_{[m}\omega_{n]}
\hat \gamma^{mn}+\frac{i}{2}f^{1/2}(\partial _mf-\omega_m
 \partial_t f)\hat \gamma^{m}\right\}(1-i\gamma^0)\biggr]\epsilon =0 \,, \\
&\biggl[{}^h\ma D_m-\omega_m\partial_t -\frac{1}{2f}(\partial_mf-\omega_m
 \partial_t f )i\gamma^0 
+\frac{f^{3/2}}{2}\left(\frac{1}{2}{{}^h\epsilon_{mn}}^{pq}\partial_{[p}\omega_{q]} +\partial_{[m}
 \omega_{n]}\right)\hat \gamma^{n}\gamma^0
\nonumber \\
&~
+\frac{i}{2f^{1/2}}\hat \gamma_m\left(kV_IX^I+\frac{i\gamma^0 \partial_t f}{2f^2}\right) 
+\left\{
-\frac{1}{4f}(\partial_nf-\omega_n \partial_t f){\hat{\gamma}_m}^{~n}-
i f^{3/2}\partial_{[m}\omega_{n]}
\right\}(1-i\gamma^0 )\biggr]\epsilon =0\,,
\end{align}
\end{widetext}
where $\hat \gamma ^m={\hat{e}}^{~m}_i \gamma^i $. 
${}^h \ma D$ and ${}^h\epsilon$ are, respectively, the Lorentz-covariant derivative and the
volume-element with respect to $h_{mn}$.   From Eqs.~(\ref{f_cube}) and 
(\ref{SUSYgauge}), we have a useful relation  
\begin{align}
kV_IX^I+ \frac 12f^{-2}{\partial_t f}=\frac{1}{2}f^2C^{IJK}H_IH_J
\left(kV_I-\frac{1}{6}\partial_t H_I\right)\,.
\end{align}
Thus,  if $\D \omega $ satisfies the anti-self duality
condition,
\begin{align}
\D \omega +\star_h \D \omega =0\,,
\label{domega}
\end{align} 
where 
$\star_h$ denotes the Hodge dual operator with respect
to the base space metric $h_{mn}$ 
and if $H_I$'s satisfy the differential equations
$\partial _t H_I=6kV_I$,
the Killing spinor equations are solved by
\begin{align}
i\gamma ^0 \epsilon &=\epsilon \label{sol_KS1} \,, \\
\epsilon &=f^{1/2}\zeta \,. 
\label{sol_KS2}
\end{align}
Here $\zeta $ is a covariantly constant Killing spinor 
with respect to the 4-dimensional metric $h_{mn}$, 
\begin{align}
{}^h\ma D_m \zeta =0\,,
\label{Killingsp_4}
\end{align}
satisfying
\begin{align}
 \hat \gamma ^{ 1 2 3 4}\zeta =\zeta \,.
\label{chiral}
\end{align}
It follows that $H_I$'s take the form,
\begin{align}
H_I(t, x^m)=6kV_I t +\bar H_I(x^m)\,, 
\label{tI}
\end{align}
where $H_I$'s are functions on the base space.

The integrability condition of Eq.~(\ref{Killingsp_4}) is
${}^hR _{mnpq}\hat\gamma^{pq}\epsilon =0$. 
From the chirality condition~(\ref{chiral}), one can find that 
$\hat\gamma_{mn}\epsilon $ is anti-self dual on the base space. 
This implies that the Riemann tensor of $h_{mn}$ is self-dual 
$\star_h (^hR_{mnpq})={}^hR_{mnpq}$. 
Hence,  the base space $(\ma B, h_{mn})$ turns out to be 
the  hyper-K\"ahler manifold whose complex structures $\mathfrak J^{(i)}$ are
anti-self-dual $\star_h\mathfrak J^{(i)}=-\mathfrak  J^{(i)}$. 
The chirality condition~(\ref{chiral}) 
is a direct consequence of $i\gamma ^0 \epsilon =\epsilon $, which is
the only projection imposed on the Killing spinor. It follows that  
the solution preserves at least half of pseudo-supersymmetries. 
If Eqs.~(\ref{tI}) and (\ref{sol_KS1}) are  satisfied,  one verifies
that the dilatino equation~(\ref{fakeKS2}) is satisfied automatically.

Let us next turn to the Maxwell equations (\ref{Maxwell_eq}). Only the
0th component is nontrivial, giving 
\begin{align}
{}^h \Delta \bar H_I=0\,,
\end{align}
where ${}^h \Delta $ is the Laplacian operator with respect to
$h_{mn}$. This equation manifests the complete linearization.

All the metric components are obtained by use of Killing spinor and
Maxwell equations under our ansatz.
We have nowhere solved the scalar and Einstein's equations so far. 
Nevertheless, these equations are automatically satisfied if the Bianchi
identities $\D F^{(I)}=0$ and Maxwell equations~(\ref{Maxwell_eq}) are
satisfied, on account of the integrability conditions for the pseudo-Killing
spinor equations.

The procedure for generating time-dependent backgrounds presented here 
was previously given in~\cite{Behrndt:2003cx}.  
It is however observed that the above metric-form is not fully general. 
According to the analysis for the de Sitter supergravity~\cite{Grover:2008jr}, the base space is
allowed to have a torsion. We expect that the general classification in this
theory is also possible following the same fashion as~\cite{Grover:2008jr}. 

\subsection{Rotating black hole in STU theory}

To be concrete, let us consider the ``STU-theory,'' 
which is defined by
the conditions such that 
$C_{123}=C_{(123)}=1$ and the other $C_{IJK}$'s vanish. In this theory,
one has three Abelian gauge fields and two unconstrained scalars.   
For simplicity, let us choose the flat space as a base space 
$(\ma B, h_{mn})$,  
\begin{align}
\D s_{\mathcal B}^2 =\D r^2+r^2 \left(\D \vartheta^2 +\sin^2\vartheta 
\D \phi_1^2+\cos ^2\vartheta \D \phi_2^2 \right)\,.
\end{align}
Then, the equation for $\omega $~(\ref{domega}) is easily solved to give
\begin{align}
\omega =\frac{J}{r^2 }\left(\sin^2\vartheta \D \phi_1 +\cos^2\vartheta
 \D\phi_2 \right)\,,
\label{omegasol}
\end{align}
where the volume form of $(\ma B, h_{mn})$ is taken as 
$\D r\we(r\D\vartheta)\we(r\sin\vartheta \D\phi_1)\we(r\cos\vartheta \D\phi_2)$
and $J$ is a constant representing the rotation of the spacetime.

In what follows we shall specialize to the case where 
each harmonic function has a point source at the origin $\propto
Q_I/r^2$.
Denoting 
\begin{align}
t_I=(6kV_I)^{-1}\,,
\end{align}
we classify the solutions into the following four cases 
depending on how many  $V_I$'s vanish~\footnote{
We shall not examine the cases in which some charges vanish $Q_I=0$
since these cases will provide nakedly singular spacetimes without
regular horizons.
}.

\bigskip\noindent (i)
$V_1=V_2=V_3=0$ for which 
\begin{align}
 H_1=1+\frac{Q_1}{r^2} \,,\qquad 
 H_2=1+\frac{Q_2}{r^2} \,, \qquad 
 H_3=1+\frac{Q_3}{r^2} \,.
\label{sol_i}
\end{align}
This is nothing but the solution in 
the ungauged true supergravity in which
the scalar field potential vanishes. The  
supersymmetric solutions have been completely 
classified in~\cite{GR,Elvang:2004ds}. 
This theory can be uplifted to 11-dimensional supergravity as described
later. The 11-dimensional solution describes the rotating
M2/M2/M2-branes preserving 1/8-supersymmetry.  
In the following, we do not elaborate this case unless otherwise stated 
since its physical properties have been widely discussed 
in the existing literature~\cite{BMPV,BLMPSV,GMT}.

\bigskip\noindent (ii)
$V_1 \ne 0 $, $V_2=V_3=0$ for which 
\begin{align}
 H_1=\frac{t}{t_1}+\frac{Q_1}{r^2} \,,\qquad 
 H_2=1+\frac{Q_2}{r^2} \,, \qquad 
 H_3=1+\frac{Q_3}{r^2} \,.
\label{sol_ii}
\end{align}
This case corresponds also to the zero potential
$V=27C^{IJK}V_IV_JX_K =0$ due to $C_{11K}=0$. 
It is notable that the potential height $V_1$ makes a contribution 
to the pseudo-Killing spinor equations~(\ref{fakeKS1}) and (\ref{fakeKS2}).  
This pseudo-supersymmetric solution can be oxidized 
to 11-dimensions, but
 the resultant spacetime is not pseudo-supersymmetric since
11-dimensional supergravity has no potential term. 
The oxidized solution is interpreted as the intersecting M2/M2/M2-branes 
in the background rotating Kasner universe.
The detail is described in section~\ref{M2M2M2}.

\bigskip\noindent (iii) $V_1, V_2 \ne 0 $, $V_3=0$ for which
\begin{align}
 H_1=\frac{t}{t_1}+\frac{Q_1}{r^2} \,,\qquad 
 H_2=\frac{t}{t_2}+\frac{Q_2}{r^2} \,, \qquad 
 H_3=1+\frac{Q_3}{r^2} \,.
\label{sol_iii}
\end{align}
These two cases (ii) and (iii) have not been discussed in~\cite{Behrndt:2003cx} although
the authors arrived at the same equation as~(\ref{tI}).

\bigskip\noindent (iv) $V_1, V_2, V_3 \ne 0 $ for which 
\begin{align}
 H_1=\frac{t}{t_1}+\frac{Q_1}{r^2} \,,\qquad 
 H_2=\frac{t}{t_2}+\frac{Q_2}{r^2} \,, \qquad 
 H_3=\frac{t}{t_3}+\frac{Q_3}{r^2} \,.
\label{sol_iv}
\end{align}
When $t_1=t_2=t_3$ and $Q_1=Q_2=Q_3$, all scalar fields are trivial.  
This case corresponds to the fake de Sitter supergravity for which the potential is constant 
$\mathfrak g^2 V=-3/(2t_1^2)$. The complete classification of
timelike class for the de Sitter supergravity was done 
in~\cite{Gutowski:2009vb,Grover:2008jr}.

Even if $t_I$'s and $Q_I$'s are not all identical, this solution inherits many
properties of that in de Sitter supergravity, irrespective of nontrivial
scalar fields $X^I$.   In fact, by a coordinate transformation
\begin{align}
 &
r' = r \left(\frac{t}{t_0}\right)^{1/2} \,, \qquad 
\ln \left(\frac{t}{t_0}\right)=\frac{t'}{t_0}+\int ^{r'}
 \frac{h_2(r')}{h_1(r')}\D  r'  \,,  
\nonumber \\
&
\phi_{1,2}= \phi_{1,2}'+\int ^{r'} h_2(r')\D  r'\,,
\end{align}
where $t_0\equiv (t_1t_2t_3)^{1/3} $ and 
\begin{align}
&
h_1( r'):=\frac{J r'^2 t_0}{H^3 r'^6-J^2} \,, \qquad
h_2( r'):=\frac{2Jr' t_0}{J^2-H^3r'^6+4r'^4 t_0^2 } \,, 
\nonumber \\
&
H^3:=\left(\frac{t_0}{t_1}+\frac{Q_1}{r'^2}\right)
\left(\frac{t_0}{t_2}+\frac{Q_2}{
 r'^2}\right)\left(\frac{t_0}{t_3}+\frac{Q_3}{r'^2}\right)\,,
\end{align}
the metric~(\ref{sol_iv}) can be brought to  the stationary form,
\begin{widetext}
\begin{align}
\D s^2 =&\frac{{r'}^2 H}{4 t_0^2}\D {t'}^2-H^{-2} 
\left[\D t'+\frac{J}{{r'}^2} \left(\sin^2\vartheta\D \phi_1'
 +\cos^2\vartheta\D \phi_2' \right)\right]^2 \nonumber \\
&+H \left[
\frac{\D {r'}^2}{1-H^3 {r'}^2/(4t_0^2)+J^2/(4t_0^2{r'}^4)}+{r'}^2 \left(\D
 \vartheta^2 +\sin^2\vartheta \D {\phi_1'}^2 +\cos^2\vartheta \D  
{\phi_2'}^2\right)\right]\,.
\label{KSsol}
\end{align}
\end{widetext}
This is asymptotically de Sitter with curvature radius 
$\ell =2t_0$~\cite{Klemm:2000vn}.

When the rotation vanishes ($\omega=0$),
these solutions reduce to the ones considered in our previous
papers~\cite{MN,MNII}, describing a spherically symmetric black hole 
in 5-dimensional FLRW universe. 
It is then expected that the present solution 
describes a rotating black hole in the expanding universe. 
To see this more concretely, 
let us consider the asymptotic limit $r\to \infty $ of the solutions.   
Let $n$ denote the number of time-dependent harmonics, i.e.,  
$n=1, 2$ and 3 are the cases (ii),  (iii) and  (iv), respectively. 
Changing to the new time slice 
\begin{align}
 \frac{\bar t}{\bar t_0}=\left(\frac{t}{t_0}\right)^{1-n/3}\,, \qquad 
\bar t_0=\frac{3t_0}{3-n}\,,
\end{align}
for $n=1,2$ and 
$\bar t=t_0 \ln (t/t_0)$  for $n=3$,  one easily finds that 
each solution (\ref{sol_i})--(\ref{sol_iv})
approaches to the 5-dimensional flat FLRW universe, 
\begin{align}
\D s_5^2=-\D \bar t^2+a^2\delta_{mn}\D x^m \D x^n  \,. 
\end{align}
Here, $\bar t$ measures the cosmic time at infinity and the scale factor
obeys
\begin{align}
a = (\bar t/\bar t_0) ^{n/[2(3-n)]}\,,
 \label{scale_factor1}
\end{align} 
for $n=1,2$ and 
\begin{align}
a= e^{\bar t/2 t_0} \,, 
\label{scale_factor2}
\end{align}
for $n=3$, 
which are respectively the same expansion law
as the stiff-matter dominant universe ($n=1$), the universe filled by
fluid with equation of state $P=-\rho/2$  
($n=2$), and the de Sitter universe with curvature radius $2t_0$ ($n=3$). 
In either case, the solution tends to be spatially 
homogeneous and isotropic in the asymptotic region 
$r\to \infty $.

On the other hand,  when one takes the limit in
which  $r$ goes to zero {\it with $t$ kept finite}, the
solution~(\ref{sol}) approaches to a deformed AdS$_2 \times$ $S^3$:
\begin{align}
\D s^2_{r\to 0}=&- \left(\frac{r^2}{\bar Q}\right)^2 \left[\D
 t+\frac{j}{r^2}(\sin^2\vartheta \D \phi_1+\cos^2\vartheta \D \phi_2) 
\right]^2 
\nonumber \\
&
+\left(\frac{\bar Q}{r^2}\right)^2 \D r^2+\bar Q \D \Omega_3^2\,,  
\label{throat}
\end{align} 
where $\bar Q\equiv (Q_1Q_2Q_3)^{1/3}$ and $\D \Omega_3^2$ 
denotes the unit line-element of $S^3$. 
This is the same as the near-horizon geometry of a BMPV black
hole~\cite{Reall2002,GMT}, 
implying that $r=0$ is a point at the tip of an infinite throat. 
Note that when all harmonics are 
time-independent, the solution reduces to the 
BMPV black hole with a degenerate horizon at $r=0$.

It is noteworthy, however,  that this metric~(\ref{throat}) does {\it not} describe 
the geometry of a neighborhood of ``would-be horizon'' since we
have fixed the time-coordinate when taking the $r\to 0$ limit. 
As pointed out in~\cite{MN,MNII} 
the null surfaces piercing the throat correspond to the infinite 
redshift ($t\to +\infty $) and blueshift ($t\to -\infty $) surfaces.     
The structures of these null surfaces can be analyzed by taking
appropriate  ``near-horizon'' limit, as we will discuss later. 
As it turns out, the horizon, if it exists, is not extremal in general, 
contrary to the na\"ive expectations from (\ref{throat}).

The reason why we consider rotating black holes in 5-dimensions 
is that rotation is compatible with
supersymmetry in 5-dimensions. In $D$-dimensions, the gravitational
attractive force and centrifugal force behave respectively 
as  $-M/r^{D-3}$ and $J^2/M^2r^2$, so that 
the balance is maintained only in
$D=5$. 
The spinning cosmological solution in the Einstein-Maxwell-axion gravity 
is obtained via dimensional reduction of a chiral null
model in 5-dimensions~\cite{Shiromizu:1999xj}.

Incidentally, let us mention the issue of the fact that 
the action involved several gauge fields.
This is a necessary price in order to obtain the finite sized horizon area.  
Just with a single gauge field, the
spacetime becomes nakedly singular unless the scalar field potential is
a pure  cosmological constant. A specific example is given in Appendix~\ref{appA}
within the framework of the Einstein-Maxwell-dilaton gravity.

\section{Physical properties of 5-dimensional rotating black holes}
\label{sec:bhs}

Let us explore the
physical properties of the solutions~(\ref{sol_i})--(\ref{sol_iv}). 
For further simplicity of our argument, we shall confine ourselves to 
the case in which all charges are identical $(Q_1=Q_2=Q_3\equiv Q>0)$
and all the potential height are the same ($t_1=t_2=t_3\equiv t_0>0$). Then,  
the metric~(\ref{SUSYmetric}) is  described in a unified way as
\begin{align}
f=
H_T^{-n/3} H_S^{-1+n/3}
\,,
\label{sol}
\end{align}
with
\begin{align}
H_T:=\frac{t}{t_0}+\frac{Q}{r^2}\,, 
\qquad 
H_S:= 1+\frac{Q}{r^2}\,, 
\label{sol_H}
\end{align}
where $ n\,(=0, 1, 2, {\rm or}~3)$ counts the number of time-dependent 
harmonics. This section is devoted to explore physical properties of 
the solution (\ref{sol}) with (\ref{sol_H}). 
Here and hereafter, the subscript ``$T$'' and ``$S$'' will be used
consistently for the
time-dependent and time-independent quantities. 
The time-dependent and static scalar fields $X_I$ are given by
\begin{align}
X_T=\frac 13\left(\frac{H_T}{H_S}\right)^{1-n/3}\,, \qquad 
X_S=\frac 13\left(\frac{H_T}{H_S}\right)^{-n/3}\,.
\label{XTXS}
\end{align}
Similarly, the gauge fields $A^{(I)}$ are 
\begin{align}
&
 A^{(T)}=H_T^{-1}\left(\D t+\frac{J}{2r^2}\sigma_3^R\right)\,,
\nonumber \\
&
A^{(S)}=H_S^{-1}\left(\D t+\frac{J}{2r^2}\sigma_3^R\right)\,.
\end{align}
The solution reduces to the BMPV solution describing an asymptotically flat
rotating black hole for $n=0$~\cite{BMPV,BLMPSV}, 
the Klemm-Sabra solution describing a rotating black hole in the de Sitter
universe for $n=3$~\cite{Klemm:2000vn}. 

Our previous solution describing a spherically symmetric 
black hole in the FLRW universe is recovered when the rotation 
vanishes $\omega =0$~\cite{MNII}. To make contact with the notation of
the reference~\cite{MNII}, let us define a canonical scalar field 
\begin{align}
\Phi = \sqrt{\frac{n (3-n)}{6}} \ln
 \left(\frac{H_T}{H_S}\right)\,,
\label{Phi}
\end{align}
and make the replacements the electromagnetic fields as 
\begin{align}
 (A^{(T)}, A^{(S)}) \to \frac{1}{\sqrt{2\pi}}(A^{(T)},
 A^{(S)})\,.
\label{em_MNII}
\end{align}
Then the solution (\ref{sol}) with (\ref{sol_H}), (\ref{Phi}) and
(\ref{em_MNII}) solves the field equations derived from the action,
\begin{align}
S_5 =&\frac{1}{2\kappa_5^2 }
\int \D ^5 x\sqrt{-g_5}\left[{}^5 R - (\nabla \Phi )^2
 -\frac{n(n-1)}{2t_0^2}e^{-\lambda_T\Phi}
\right.\nonumber \\& \left.
-
\sum_{A=T,S}n_A e^{\lambda_A \Phi }
F^{(A)}_{\mu\nu }F^{(A)\mu \nu }
+2\epsilon^{\mu\nu\rho\sigma\tau }
A_\mu^{(T)}F_{\nu\rho }^{(S)}F_{\sigma\tau}^{(S)}
\right]\,,
\label{5Deffaction}
\end{align}
where $n_T=3-n_S=n$ and
\begin{align}
\lambda_T =2\sqrt{\frac{2n_S}{3n_T}}\,,\qquad 
\lambda_S =-2\sqrt{\frac{2n_T}{3n_S}} \,,
\end{align}
which is the $D=5$ action considered 
in~\cite{MNII} when the Chern-Simons term does not contribute, i.e.,
there is no rotation.

When the theory is motivated by supergravity,  
the parameter $n$ takes an integer value.
We should stress that even if $n$ is not an integer,
the aforementioned metric~(\ref{SUSYmetric}) 
with (\ref{sol}) and (\ref{sol_H}) is still an exact solution of
the Einstein-Maxwell-scalar system, in which we have two ${\rm U}(1)$
fields coupled to the scalar field with an Liouville-type exponential 
potential~(\ref{5Deffaction}). 
The solution with non-integral values of $0<n<2$ is qualitatively similar to
the one with $n=1$. (The case $2<n<3$ has no representative 
in this paper.) 
The geometrical properties with $n=1$ discussed in what follows
are also applied to the solution with $0<n<2$.

\subsection{Symmetries}

At first sight, one might expect that the metric admits 
${\rm U}(1)\times {\rm U}(1)$ spatial symmetries generated by 
$\partial/\partial \phi_1$ and $\partial/\partial \phi_2$. 
In order to see that the solution indeed admits much larger symmetry, 
let us introduce the Euler angles 
$(\theta , \phi, \psi)$ by
\begin{align}
\theta =2\vartheta \,,\qquad 
\phi=\phi_2-\phi_1\,,\qquad
\psi=\phi_2+\phi_1\,,
\end{align}
which take ranges in 
$0\le \theta\le \pi$, 
$0\le \phi\le 2\pi$ and
$0\le \psi\le 4\pi$.
In terms of above coordinates, 
the left-invariant one-forms $\sigma^R_i$
($i=1,2,3$) on ${\rm SU}(2)\simeq S^3$ 
are given by
\begin{align}
\sigma_1^R &=-\sin\psi \D\theta+\cos \psi \sin\theta \D \phi \,,\\
\sigma_2^R &=\cos\psi \D \theta +\sin\psi \sin \theta \D \phi \,,\\
\sigma_3^R &=\D \psi+\cos \theta \D \phi \,.
\end{align}
These one-forms satisfy
\begin{align}
\D \Omega_3^2=\frac 14\sum_i(\sigma_i^R)^2\,,
\qquad
\D\sigma_i^R=\frac 12\sum_k\epsilon_{ijk}\sigma_j^R\wedge\sigma_k^R\,.
\end{align}
The right-invariant vector fields $\xi ^L_i$ 
are the spacetime Killing fields. They are given by
\begin{align}
\xi^L_1&=-\frac{\cos\phi }{\sin \theta }\partial_\psi+\sin\phi 
\partial_\theta +\cot \theta \cos\phi \partial_\phi \,,\\
\xi^L_2&=\frac{\sin\phi }{\sin\theta }\partial_\psi+\cos\phi
 \partial_\theta -\cot\theta \sin \phi \partial_\phi \,,\\
\xi^L_3&=\partial_\phi \,,
\end{align}
which are the generators of the left transformations of ${\rm SU}(2)$.
These Killing vectors satisfy
\begin{align}
&
\mas L_{\xi_i^L} \sigma_j^R=0 \,, \qquad
[\xi_i^L, \xi_j^L]=\sum_k\epsilon_{ijk}\xi_k^L\,, 
\nonumber \\
&
\left(\frac{\partial }{\partial \Omega_3}\right)^2 =
4 \sum_i\xi_i^L \xi_i^L\,.
\end{align}
Besides these, 
there exists an additional U$(1)$-Killing field
\begin{align}
\xi_3^R&=\partial_\psi \,.
\end{align}
The orbits of $\xi_3^R$ are the fibres of Hopf fibration of $S^3$.
It follows that the metric is invariant under the 
action of ${\rm U}(2)\simeq {\rm SU}(2)\times {\rm U}(1)$, 
acting on the 3-dimensional orbits which are spacelike at infinity. 
Thus, the metric is expressed as
\begin{align}
\D s^2=-f ^2 \left(\D t+\frac{J}{2r^2}\sigma_3^R\right)^2
+f^{-1}
\left(
\D r^2+r^2\D \Omega_3^2 \right)\,.
\label{metirc_Euler}
\end{align}

As discussed in~\cite{GibbonsBMPV}, 
the metric with ${\rm U}(2)$-symmetry admits a {\it reducible} Killing tensor
\begin{align}
\nabla _{(\mu }K_{\nu \rho )}=0\,, \qquad 
K^{\mu \nu }=\sum_i (\xi_i^L)^\mu (\xi^L_i)^\nu \,, 
\label{Killing_tensor}
\end{align}
which enables us to separate angular variables for the 
geodesic motion and scalar field equation. 
It should be remarked, however, that the solution does not admit a 
timelike Killing field, so that the geodesic motion is not 
immediately solved.

\subsection{Singularities}

One can immediately find that the scalar fields $X_I$~(\ref{XTXS}) blow up at
\begin{align}
t=t_s(r):=-\frac{t_0Q}{r^2} \quad {\rm and } \quad
r^2=-Q\,.
\label{sing}
\end{align}
Straightforward calculations reveal that all the curvature invariants are divergent 
at these spacetime points, i.e., they are 
spacetime curvature singularities.  For example, the Ricci scalar
curvature is given by
\begin{widetext}
\begin{align}
{}^5 R= \frac{f^4 }{6r^8 H_T^2}\Biggl[&\frac{2n(3n-4)r^8}{t_0^2 f^6}
+J^2 \left(24 H_T^2+\frac{n(2-n)r^2}{t_0^2f^3}\right)
\nonumber \\ &
-4 Q^2r^2
 H_T^nH_S^{1-n}\left\{2(nH_S^2+(3-n)H_T^2)-(nH_S+(3-n)H_T)^2\right\}
\Biggr]\,,
\end{align}
\end{widetext}
which diverges at the above spacetime points, as expected.

Note that the $t=0$ surface and the surface $r=0$ with 
$t$ kept finite are not the curvature singularities, where the curvature
invariants are bounded. Hence, 
the big-bang singularity at $t=0$ is completely smoothed out due to
electromagnetic charges.  As in the case (i), the surface 
$r=0$ is a plausible candidate of event horizon.

\subsection{Closed timelike curves}

Since  the vector field $\xi_3^R=\partial_\psi$ 
generates closed orbits of the period
$4\pi$, there appear closed timelike curves if an orbit of $\xi_3^R$
becomes timelike. 
Rewrite the metric~(\ref{metirc_Euler}) as
\begin{align}
\D s^2 =&-\frac{f^2}{\Delta_L}\D t^2 +\frac{\D
 r^2}{f}
+\frac{r^2}{4f}
\biggl[(\sigma_1^R)^2+(\sigma_2^R)^2
 \nonumber \\ &  +\Delta_L \left(\sigma_3^R-
\frac{2 J f^3 }{r^4\Delta_L}\D t\right)^2\biggl]\,,
\label{metric_CP}
\end{align}
where
\begin{align}
 \Delta_L:=1-\frac{J^2f^3}{r^6}\,.
\label{Delta_L}
\end{align}
Inspecting 
\begin{align}
g (\xi_3^R, \xi_3^R)=\frac{f^2}{4}\left(
H_T^nH_S^{3-n}-\frac{J^2}{r^6}
\right)\,,
\end{align}
we can see that the 
first term on the right hand side vanishes
at the singularities. It follows that 
the Hopf fibres become timelike, i.e., 
closed timelike curves inevitably emerge in the vicinity of
singularities for all values of $J (\ne 0)$. 
$\Delta_L=0$ defines the velocity of light surface
(VLS), where closed causal curves appear for $\Delta _L<0$. 
For $n=0$ (the BMPV spacetime without time-dependence), 
the VLS is located at 
$r^2 =J^{2/3}-Q$ which is inside the horizon for the small rotation 
$J^{2/3}<Q$, otherwise it is outside the horizon.

For $n\ne 0$, the VLS has the time-dependent profile 
\begin{align}
t_{\rm VLS}(r):= \frac{t_0}{r^2}\left[\left\{\frac{J^2}
{(r^2+Q)^{3-n}}\right\}^{1/n}-Q\right]\,.
\label{VLS}
\end{align}
Since $S^3$ is ${\rm U}(1)$-fibration over $S^2$, 
one can introduce the radius of $S^2$ by 
\begin{align}
R= |r |f^{-1/2}\,.
\end{align} 
In terms of $R$, the VLS is positioned at  the constant radius, 
\begin{align}
R_{\rm L}=J^{1/3}\,.
\end{align}
We shall say the region $R<R_L$ ($R>R_L$) inside (outside) the VLS. 
Inside the VLS,
 $\xi_3^R$ is pointing into the future direction
for $J>0$ and into the past one for $J<0$. 
It is obvious that the singularity $t_s(r)$ exists for $r>0$. 
As we will see in the next subsection,
a horizon is positioned at $r=0$ (with
$t=\infty$), so
that these closed timelike curves yield naked time machines--the 
causally anomalous region that is not hidden behind the
 event horizon--for every
choice of parameters. 
Since the spacetime is simply connected, these causal pathologies cannot
be circumvented by extending to a universal covering space.  
Hence the fake supersymmetry fails to get rid of causal pathologies, 
as occurred for the present BPS rotating solutions.

Figure~\ref{fig:VLS} plots the typical behaviors of VLS. 
When the angular momentum $J$ is smaller than the critical 
value $Q^{2/3}$, 
$t_{\rm VLS}(r)<0$ is satisfied and 
$t_{\rm VLS}(r)\to -\infty $
as $r\to 0$. On the other hand, 
when the angular momentum is larger than $Q^{3/2}$, 
$t_{\rm VLS}(r) \to +\infty $ as $r\to 0$ and  for $n=3$  
$t_{\rm VLS}(r) $ is always positive, 
whereas it is negative at large value of $r$ for $n=1,2$.  

Using the radius $R$ one finds that the singularities~(\ref{sing}) are
both central $R=0$. So the VLS completely encloses the spacetime
singularities.

In the neighborhood of the VLS,  
$(t-t_{\rm VLS}(r))/t_0\ll 1$, one finds  
$f^{-3}\simeq J^2/r^6$. Hence the neighborhood of the VLS
in the present spacetime may be approximated  by that 
in the near-horizon geometry of the 
BMPV black hole~(\ref{throat}) 
with $J^2=Q^3$. 
In this case,  $\xi_3^R =\partial/\partial \psi $ 
becomes a hypersurface-orthogonal null Killing vector. Moreover $\psi $ 
corresponds to the affine parameter of the null geodesics 
$(\xi_3^R)^\nu \nabla_\nu (\xi_3^R)^\mu =0$, so that  the spacetime 
describes the
plane-fronted wave (not the plane-fronted wave with parallel rays
(pp-wave) since $\xi_3^R $ is not covariantly
constant $\nabla_\mu (\xi_3^R)^\nu  \ne 0$)~\cite{Stephani:2003tm}.  
We can expect that properties of the VLS in the present spacetime
is captured by that in the near-horizon geometry of the BMPV black hole
with $J=Q^{3/2}$.

\begin{widetext}
\begin{center}
\begin{figure}[h]
\includegraphics[width=14cm]{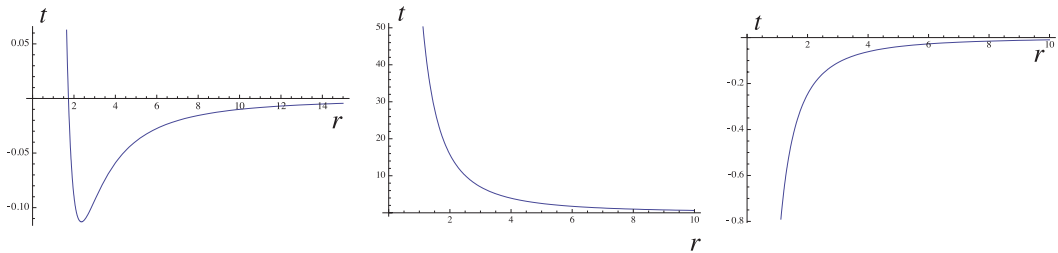}
\caption{Plots of velocity of light surface $t_{\rm VLS}$ 
against $r$ for 
$n=1, 2$ with $J^{2/3}>Q$ (left), for 
$n=3$ with $J^{2/3}>Q$ (middle) and for $J^{2/3}<Q$. }
\label{fig:VLS}
\end{figure}
\end{center}
\end{widetext}

\subsection{Scaling limit}

Since the event horizon of a black hole is  a global concept, 
it is a difficult task to identify its locus especially in a
time-dependent spacetime. Following the previous papers we 
shall argue the ``near-horizon geometry'' of the present metric and 
demonstrate that the null surface of the event-horizon candidate
is described by a Killing horizon. By solving null geodesics
numerically, we can verify that when $J<Q^{2/3}$ these Killing horizons are indeed event horizons
in the original spacetimes. (The reason of the restriction $J<Q^{2/3}$
will be discussed later.)

For convenience, we define dimensionless parameters
\begin{align}
 \tau:=\frac{t_0}{Q^{1/2}}\,,\qquad 
 j :=\frac{J}{Q^{3/2}}\,,
\end{align}
and denote dimensionless variables (normalized by $Q$) with tilde,  
e.g., $\ti x^\mu :=Q^{-1/2} x^\mu $. Then we can 
work with the dimensionless metric, 
\begin{align}
 \D \tilde s^2 =&- f^2
\left(\tau \D \tilde t+\frac{j}{2\tilde r^2 }\sigma_3^R \right)^2
+f^{-1}\left(\D \tilde r^2+\tilde r^2 \D \Omega_3^2\right)
 \,, \nonumber \\
f=& \tilde r^2 \left(\tilde t\tilde r^2+1
 \right)^{-n/3}
\left(\tilde r^2+1\right)^{(n-3)/3}\,. 
\end{align} 
The parameter $j$ is the reduced angular momentum and $\tau $
denotes the ratio of energy densities of the scalar fields and the Maxwell
fields evaluated on the horizon, respectively~\cite{MN,MNII}.  
To simplify the notation, we shall omit tilde in the following.

We have seen in Eq.~(\ref{throat}) that the surface $r\to 0$
 with $t$ being finite 
corresponds to the throat infinity. Hence the null surfaces 
``intersecting'' at the throat should be a candidate of 
future and past horizons. These surfaces are described by the 
infinite redshift and blueshift surfaces, respectively. 
We shall focus on the geometry of the very neighborhood 
of these  horizon candidates. 
The only well-defined ``near-horizon'' limit is given by
\begin{align}
t~\to~\frac{t}{\epsilon ^2}\,, \qquad r~\to ~\epsilon r \,, 
\qquad \epsilon~\to ~ 0\,,
\label{NHlimit}
\end{align}
under which the metric is free from the scaling parameter $\epsilon $.  
The above scaling limit gives rise to the near-horizon geometry,
if a horizon exists, the metric of which is given by 
\begin{align}
\D  s^2_{\rm NH} =&- r^4\left( t r^2+1\right)^{-2n/3}
\left(\tau \D  t+\frac{j}{2 r^2}\sigma_3^R\right)^2 \nonumber \\
&+ r^{-2}
\left( t r^2+1\right)^{n/3}\left(\D  r^2+ r^2\D \Omega_3^2
\right)\,.
\label{NHmetric}
\end{align}
The scalar and gauge fields are also well-defined and given by
\begin{align}
 X_T=\frac 13\left(tr^2+1\right)^{1-n/3}\,, 
\qquad 
 X_S=\frac 13\left(tr^2+1\right)^{-n/3}\,,
\label{NHscalar}
\end{align}
and
\begin{align}
&
A^{(T)}=r^2 \left(tr^2+1\right)^{-1}\left(\tau \D
 t+\frac{j}{2r^2}\sigma_3^R\right)\,, 
\nonumber \\
&
A^{(S)}=r^2 \left(\tau \D
 t+\frac{j}{2r^2}\sigma_3^R\right)\,.
\end{align}
This spacetime is  pseudo-supersymmetric in its own right since  
it admits a nonvanishing
Killing spinor of the form (\ref{sol_KS1}) and (\ref{sol_KS2}) 
with $f=r^2(tr^2+1)^{-n/3}$.

The above near-horizon metric~(\ref{NHmetric}) is still time-evolving and
spatially inhomogeneous. Nevertheless,  
as a consequence of the scaling limit~(\ref{NHlimit}), 
the near-horizon metric~(\ref{NHmetric}) admits a 
Killing vector
\begin{align}
\xi^\mu = t\left(\frac{\partial }{\partial  t}\right)^\mu 
-\frac{ r}{2}\left(\frac{\partial }{\partial  r}\right)^\mu \,.
\label{xi}
\end{align}
It is then convenient to take $\xi^\mu $ to be a coordinate vector so that the 
metric is independent of that coordinate. 
A possible coordinate choice ($T, R, \psi'$) is given by 
\begin{align}
T =&\ln | t|
+\int ^R \frac{6 R^{6/n-1}(R^6-j^2)\D R}{n (R^{6/n}-1)\Delta }\,,
\nonumber \\
 R=&( t r^2+1)^{n/6}\,, 
\nonumber \\
\psi' =&\psi+\int^R \frac{12 j\tau  R^{n/6-1}}{n \Delta }\D R\,,
\label{coord_change}
\end{align}
where
\begin{align}
\Delta :=&4 R^4 F( R)+j^2 \,, \nonumber \\
F( R):= &\tau^2  R^{-4} ( R^{6/n}-1)^2 -\frac 14 R^2 \,,
\end{align}
In this new coordinate system, the Killing field is simply given by
$\xi^\mu =(\partial/\partial  T)^\mu $, as we desired. 
After some algebra the near-horizon metric~(\ref{NHmetric}) is cast
into an apparently stationary form, 
\begin{align}
\D  s^2_{\rm NH}=&
-F( R)\left[\D  T+\frac{ j\tau  ( R^{6/n}-1)}{2 R^4 F(R)}
{\sigma _3'}^R\right]^2 +\frac{j^2  R^2({\sigma_3'}^R)^2}{16F( R)}
\nonumber \\ &
+\frac{36\tau^2  R^{12/n}\D  R^2 }{n^2 \Delta }
 +\frac{ R^2}4 \left[(\sigma_1^R)^2+(\sigma_2^R)^2+({\sigma_3'}^R)^2\right]\,. 
\label{NHmetric2}
\end{align}
Here, ${\sigma_3'}^R=\D \psi '+\cos\theta \D \phi$.  
Although its asymptotic structure is highly nontrivial, it is easy to recognize
that this spacetime has Killing horizons (if any) at $\Delta =0$. 
The Killing horizon is generated by a linear combination of stationary and
angular Killing vectors, 
\begin{align}
\zeta  =\frac{\partial }{\partial T}+2\Omega_h \frac{\partial
 }{\partial \psi'}\,, 
\label{generator}
\end{align} 
where
\begin{align}
\Omega_h=\left.\frac{j}{2 \sqrt{R ^6-j^2 }}\right|_{\rm horizon}\,. 
\end{align}
Here, $\Omega _h$ is the angular velocity of the horizon
(associated with  
$2\partial/\partial \psi'=\partial/\partial\phi_2+\partial/\partial{\phi_1}$). 
The horizon angular velocity $\Omega_h$  is
constant anywhere on the horizon, which is a generic feature of
a Killing horizon~\cite{Carter}. 
Contrary to (truly) supersymmetric black holes, the angular velocity
of the horizon is nonvanishing, i.e., the horizon is {\it rotating}.
In other words, the generator of the event horizon of a supersymmetric
black hole is tangent to the stationary Killing field at infinity.   
Equation~(\ref{generator}) shows that $\partial/\partial t $ is not
the generator of the event horizon.  This is a distinguished property 
not shared by the BPS black holes.

Since $\Delta $ fails to have a double root in general, it follows that
the horizon is not extremal unless parameters $(\tau, j)$ are
fine-tuned. The reason of the appearance of the ``throat'' geometry at $r\to 0$ lies
in the fact that $(t, r)$-coordinates cover the ``white-hole region''
as well as the outside region of a black hole (see Figure~5 in~\cite{MNII}).

Equations~(\ref{NHscalar}) and~(\ref{coord_change}) imply 
that the values of scalar fields $X_I$ on the horizon
 are determined by the horizon radius,
which is expressed in terms of the charge $Q$, (inverse of) potential
height $t_0$ and the angular momentum $j$. This situation is closely
analogous to the attractor mechanism~\cite{attractor}, according to
which the values of scalar fields on the horizon are expressed by
charges and  independent of the asymptotic values of the scalar 
fields at infinity. But as it stands 
it appears hard to say whether such a mechanism always works 
in the time-dependent case.

In the following subsections we shall clarify various physical 
features of the near-horizon metric~(\ref{NHmetric2}).

\subsubsection{Horizons}

The loci of Killing horizons $\Delta=0$ can be classified according to the 
values of $\tau $ and $j^2$. We shall say ``under-rotating'' when the 
spacetime~(\ref{NHmetric}) admits horizons. Otherwise it is said to be 
``over-rotating.'' The quantity $R_-$ will be consistently used 
when $\Delta >0$ for $R<R_-$. 



\bigskip\noindent
(i) $n=1$. 
When the angular momentum parameter $|j|$ is less
       than the critical value $j_{(1)}$, i.e.,
\begin{align}
 j^2< j_{(1)}^2:=\frac{1+16\tau^2 }{16\tau^2 }\,,
\end{align}
the near-horizon spacetime admits two horizons,
\begin{align}
 R_\pm^6 =\frac{1+8\tau^2 \pm \sqrt{1+16\tau ^2 (1-j^2)}}{8\tau^2 }\,. 
\label{Rpm_1}
\end{align} 
For the over-rotating case $j^2>j_{(1)}^2$,  
there exist no horizons.  
We find the similar results for the case of non-integer values of $n<2$.


\bigskip\noindent
(ii) $n=2$. This case is further categorized into the following three cases.
 
\begin{enumerate}
 \item $0<\tau<1/2$. For any values of $j$, a single horizon occurs at
 \begin{align}
  R_-^3 =\frac{\sqrt{4\tau^2 +(1-4\tau^2 )j^2}-4\tau^2 }{1-4\tau^2 }\,. 
\label{Rm2}
 \end{align} 
\item $\tau =1/2$. For any values of $j$, a single horizon occurs at
\begin{align}
R_-^3 =\frac{1+j^2}{2}\,. 
\label{Rm3}
\end{align}

\item $\tau >1/2$.  When the angular momentum parameter $|j|$ is less
       than the critical value $j_{(2)}$, i.e.,
\begin{align}
 j^2< j_{(2)}^2:=\frac{4\tau^2 }{4\tau^2-1 }\,,
\end{align}
two horizons exist at 
\begin{align}
 R_\pm^3 =\frac{4\tau^2 \pm \sqrt{4\tau^2
 -(4\tau^2-1)j^2} }{4\tau^2-1 }\,.
\label{Rm4}
\end{align}
For the over-rotating case $j^2>j^2_{(2)}$, 
no horizons develop.

\end{enumerate}


\bigskip\noindent
(iii) $n=3$. 
In this case, 
the metric~(\ref{NHmetric2}) is not the ``near-horizon''
geometry, but is the original metric itself written in the stationary 
coordinates. This metric describes a charged rotating black hole in de Sitter
space derived by Klemm and Sabra~\cite{Klemm:2000vn}.  
Let us discuss its horizon structure in detail.

There exists at least one horizon corresponding to the cosmological
horizon.  For $\tau \le \sqrt{3/2}$ there appears only a cosmological horizon
$R_c$. For $\tau>\sqrt{3/2}$, the number of horizons depend on the value of $j^2$. 
Three distinct horizons ($R_-<R_+<R_c$) exist for 
$j^2_{(3)-}<j^2<j^2_{(3)+}$, where
\begin{align}
j^2_{(3)\pm}&:=\frac{4\tau^2 }{27}\left[\pm 
8\sqrt 2\tau (2\tau^2-3)^{3/2}
\nonumber \right.
\\
&
\left.~~~~~~~~~~
-32\tau^4
+9(8\tau^2-3)\right]\,.
\end{align}
For $j^2 =j^2_{(3)+}$, inner and outer black-hole horizons  are
      degenerate.  While,  
for $j^2 =j^2_{(3)-}$ the outer black-hole horizon   and the
      cosmological horizon are degenerate. 
$j^2_{(3)-}$ takes real positive values for $\sqrt{3/2}<\tau <3\sqrt{3}/4$,  
otherwise the inner horizon does not exist.

\bigskip
A simple calculation reveals that the spacetime~(\ref{NHmetric}) is
regular on and outside the Killing horizon (if any). Only the existing
curvature singularity is at $R=0$. It is almost clear to construct the
local coordinate systems that pass through the Killing horizon 
$\Delta =0$.

In hindsight, 
we can understand why the horizon in the present spacetime 
is not extremal as follows. 
In the case of the time-independent (truly) BPS solutions such as a BMPV
black hole,  the Killing horizon lies at $f=0$ 
since $V^\mu =(\partial/\partial t)^\mu $ is an everywhere causal 
Killing field constructed
by a Killing spinor $\epsilon $ as 
$V^\mu =i \bar \epsilon \gamma^\mu\epsilon$ (see~\cite{GGHPR}).   
For the present time-dependent pseudo-supersymmetric black hole, 
on the other hand, 
the vector field $V^\mu =(\partial/\partial t)^\mu$ is not the 
Killing-horizon generator: the horizon is generated by 
$
\xi ^\mu =t(\partial/\partial t)^\mu -(r/2)(\partial/\partial r)^\mu 
+\Omega_h(\partial/\partial \psi)^\mu 
$ 
given in Eq.~(\ref{generator}). 
The vector field  $V^\mu$ does not give rise to any (asymptotic)
symmetry.

Physically speaking, the degeneracy of the horizon is broken by
introducing of the time-dependent scalar fields (which do not contribute
to the total mass when the spacetime is stationary) or the positive 
cosmological constant.  These ingredients destroy the fine balance  
between the mass energy and the charges.
When  the rotation is also added the centrifugal force
gives a negative contribution to the mass energy $M\to M-J^2$--which
takes place only in $D=5$ as discussed before--thus it exceeds the
extremal threshold value if the rotation becomes too large. 

\bigskip

%

\subsubsection{Ergoregion}

An obvious major difference from our previous non-rotating
solutions~\cite{MN,MNII} is that  
the near-horizon metric possesses the ergosurface at $F(R)=0$.
Since $\Delta >4R^4 F(R)$, the ergosurface  
lies strictly outside the horizon, contrary to the 4-dimensional 
Kerr black hole for which the ergosurface touches the 
horizon at the rotation axis.

When the rotating vanishes ($j=0$), 
the roots of $F=0$ correspond to the loci of horizons~\cite{MN,MNII}. 
Since $\Delta =0$ reduces to $F(R)=0$ when $j=0$, 
the explicit expression of the ergosphere is given by setting $j=0$ of the 
horizon radius. For $n=1,2$ they are given by Eqs.~(\ref{Rpm_1}),
(\ref{Rm2}), (\ref{Rm3}) and (\ref{Rm4}) with $j=0$.  
Note, however,  that since  the asymptotic structures are quite
peculiar when $n=1,2$, there may arise an ambiguity concerning the definition
of the energy~\footnote{For example, the AdS spacetime has various
Killing fields which are timelike outside the AdS degenerate horizon. 
The definition of energy in asymptotically AdS spacetime 
is different depending on which Killing field is chosen.
}.
It may therefore equivocal whether $R_{\rm erg}$ has a definitive meaning 
in the $n=1,2$ cases.

When $n=3$ the asymptotic region is described by de Sitter space, 
so that we can use the standard time translation with respect to the 
observer at the cosmological horizon to define the energy. 
Hence the notion of ergoregion is meaningful in this sense. 
There exist three distinct roots, 
$ R_{\rm erg, -}< R_{\rm erg, +}< R_{\rm erg, c}$, for 
$\tau >\tau _{\rm cr}:=3\sqrt 3/4$, two roots 
for $\tau=\tau_{\rm cr}$ and a single root $ R_{\rm erg, -}$
for $\tau <\tau _{\rm cr}$.

The ergoregion does not arise for the  
supersymmetric black hole, which inevitably forbids the ergoregion
inside which the stationary Killing field becomes spacelike. 
The ergoregion is intrinsic to a rotating black hole and allows 
particles to have a negative energy. This means that the rotation energy
of a black hole can be subtracted via the Penrose process and the
superradiant scattering process. We shall demonstrate in
Appendix~\ref{sec:superradiance} that this is
indeed the case for the $n=3$ Klemm-Sabra solution.

\subsubsection{Closed timelike curves}

Write the near-horizon metric~(\ref{NHmetric2}) as
\begin{widetext}
\begin{align}
\D s_{\rm NH}^2 =- \frac{\Delta }{4 R^4\varDelta_L }\D T^2 
+\frac{36\tau^2 R^{12/n}\D R^2}{n^2\Delta }+\frac{R^2}{4}\left[
(\sigma_1^R)^2 +(\sigma_2^R)^2 +\varDelta_L\left({\sigma_3'}^R-
\frac{2j\tau (R^{6/n}-1)}{R^6\varDelta_L}\D T\right)^2 \right]\,,
\end{align}
\end{widetext}
where we have also used  $\varDelta_L$ as the near-horizon 
limit of (\ref{Delta_L}):
\begin{align}
 \varDelta_L= 1-\frac{j^2}{ R^6}\,.
\label{Delta_LNH}
\end{align}
Consequently 
the Hopf fibres become timelike inside the VLS ($\varDelta_L<0$), viz, 
the near-horizon metric~(\ref{NHmetric2}) is also causally unsound. 
In terms of $\varDelta_L$, $\Delta $ is 
\begin{align}
 \Delta =4\tau^2 (R^{6/n}-1)^2 -R^6\varDelta_L \,.
\end{align} 
It follows that the event horizon ($\Delta =0$) is outside the 
VLS~\cite{Matsuno:2007ts}.  
Hence {\it the causality violating region
is always hidden behind the horizon} ($ R_L< R_+$) in the near-horizon 
geometry~(\ref{NHmetric2})~\footnote{
Since $R_-^6-R_L^6=4\tau ^2(R_-^2-1)^2>0$,  $R_L<R_-$ also holds.}. 
On the other hand, it is naked in the 
over-rotating case where the horizon does not exist. 
This should be contrasted with the BMPV or asymptotically AdS
($\mathfrak g\in \mathbb R$)
Klemm-Sabra black hole. In the former case   
the VLS is outside the event horizon if the angular momentum is 
large $J>Q^{2/3}$. 
In the latter case a naked time machine 
inevitably appears outside the event horizon.
However, it  allows no geodesics
to  penetrate, so that the horizon exterior is geodesically complete.  
In the present case the area of the horizon is given by
\begin{align}
 {\rm Area} &=2\pi ^2 R^3\left.\sqrt{\varDelta_L}\right|_{\rm horizon}
=4\pi^2\tau (R_+^{6/n}-1) 
\,.
\end{align}
which always makes sense contrary to the BMPV or the asymptotically AdS
Klemm-Sabra black hole: the latter two spacetimes have 
an ``imaginary horizon area'' in the
over-rotating case. These formal horizons in the over-rotating case
are ``repulsons'' into which no freely falling orbits 
penetrate~(see e.g.,~\cite{GibbonsBMPV,Caldarelli:2001iq,Herdeiro:2000ap,Dyson:2006ia}).

\subsubsection{Geodesic motions}

It is illustrative to consider geodesics in the near-horizon
metric~(\ref{NHmetric2}). For $n=3$, the following analysis  
yields the geodesic motion in the exact Klemm-Sabra geometry, not
restricted in the neighborhood of its horizons. 
The particle motion in asymptotically AdS Klemm-Sabra solution 
($\mathfrak g \in \mathbb R $) was
previously examined in~\cite{Caldarelli:2001iq}. 
Although the 
behavior of the particle motion in asymptotically 
de Sitter case is of course considerably different from that case, 
the technical method is similar. 
The analysis in this subsection unveils that the horizon can be reached
within a finite affine time from outside.

The Hamilton-Jacobi equation in the near-horizon 
geometry~(\ref{NHmetric2}) reads  
\begin{align}
- 
\frac{\partial S}{\partial \lambda }
= 
\frac 12g^{\mu\nu}_{\rm NH}\left(\frac{\partial S}{\partial x^\mu}\right)
\left(\frac{\partial S}{\partial x^\nu}\right)\,,  
\end{align}
where the right-hand side of this equation defines a geodesic
Hamiltonian and $\lambda $ is an affine parameter.
Assume the separable form of Hamilton's principal
function, 
\begin{align}
 S=& \frac 12m^2\lambda -E  T +L_L\phi+L_R \psi' 
+S_R ( R)+S_\theta (\theta)\,,
\label{pri_fun}
\end{align}
where $E, L_R, L_L$ and $m$ are constants of motion corresponding to
energy, right-rotation, left-rotation and rest mass of a
particle. 
Since the near-horizon metric keeps the ${\rm U}(2)$-symmetry, 
there exists a reducible Killing tensor of the
form~(\ref{Killing_tensor}), which reads
in the coordinates (\ref{NHmetric2}) as
\begin{align}
K_{\mu\nu }\D x^\mu \D x^\nu =&\left[
\frac{j\tau }{2R^4}(R^{6/n}-1)\D T-\frac{R^2\varDelta _L}{4}
({\sigma'_3}^R)\right]^2 \nonumber \\
&
+\frac{R^4}{16}\left[(\sigma_1^R)^2+(\sigma_2^R)^2 \right]\,.
\\
\nonumber 
\end{align}
Accordingly, besides obvious constants of motion ($E, L_R,L_L, m$) 
generated by Killing vectors, 
we have an additional integration constant $L^2$ with dimensions of 
angular momentum squared such that  
\begin{align}
L^2:= \sum _i(\xi^R_i S)^2 \,. 
\label{Lsq}
\end{align}

This constant of motion enables us to separate the variables as 
\begin{align}
\left(\frac{\D }{\D \theta }S_\theta
\right)^2+\frac{1}{\sin ^2 \theta }
\left(L_L^2+L_R^2-2\cos\theta L_RL_L\right)=L^2\,.
\label{Lsq2}
\end{align}
The constant $L^2$ represents the left and right Casimir invariant
of the ${\rm SU}(2)$ subgroup of ${\rm SO}(4)$ rotation group. 
These two Casimirs turn out to be the same for the scalar representation. 
It follows that the particle motion and the scalar-field equation are 
Liouville-integrable.  
The governing equations are obtainable by differentiating the 
principal function~(\ref{pri_fun}) by corresponding constants of
motion. Using the relation for angular variable~(\ref{Lsq2}), 
we obtain a set of useful 1st-order equations   
\begin{widetext}
\begin{align}
 \frac{\D  R}{\D \lambda }&=\pm
\frac{n\sqrt {\Delta }}{6\tau  R^{6/n}}
\left[\frac{E}{F}-m^2-\frac{4 (L^2-L_R^2)}{ R^2}
-\frac{16 R^2 F}
{\Delta }\left(L_R+\frac{j\tau ( R^{6/n}-1)E}{2 R^4
 F}\right)^2\right]^{1/2}\,,\label{NH_radeq}\\
\frac{\D \theta }{\D \lambda } &= \pm \frac{4}{R^2}\left[
L^2-\frac{1}{\sin ^2 \theta }
\left(L_L^2+L_R^2-2\cos\theta L_RL_L\right)
\right]^{1/2} \,,\\
\frac{\D  T}{\D \lambda }&=\frac{4 R^4 \varDelta_L}{\Delta
 }E -\frac{8j\tau (
 R^{6/n}-1)L_R}{\Delta  R^2}\,,\label{NH_timeeq}\\
\frac{\D \phi}{\D \lambda }&=\frac{4}{ R^2 \sin^2\theta }
(L_L-L_R\cos \theta )\,,\label{NH_phieq}\\
\frac{\D \psi' }{\D \lambda }&=\frac{4(L_R-L_L\cos\theta )}{
 R^2\sin^2\theta }
-\frac{4}{R^2\Delta }\left[j^2 L_R-2j\tau (R^{6/n}-1)E\right]\,.
\label{NH_psieq}
\end{align}
\end{widetext}
If there are no angular momenta of a particle ($L=L_R=L_L=0$),
one sees that there is no motion in directions
$\theta $ and $\phi$, but there is a nonvanishing motion along $\psi'$,
encoding the frame dragging due to the black-hole rotation.

By virtue of high degree of symmetries, the problem reduces  to the one
dimensional radial equation~(\ref{NH_radeq}), which is arranged to give 
\begin{align}
\left(\frac{\D R}{\D \lambda }\right)^2=&
\frac{n^2}{9 \tau^2 R^{4(3/n-1)}}\biggl[
\varDelta_L (E-2\Omega L_R)^2
\nonumber \\ & 
-\frac{j^2L_R^2\Delta }{R^{12}\varDelta_L}-\Delta 
\left(\frac{L^2}{R^6}+\frac{m^2}{4R^4}\right)\biggl]\,,\label{NH_radeq2}\\
=&
\frac{n^2\varDelta _L}{9 \tau^2
 R^{4(3/n-1)}}(E-V^+)(E-V^-)\,,\label{NH_radeq3}
\end{align}
where $\Omega $ and $V^\pm$ are the angular velocity of a locally
nonrotating observer and the effective potentials, 
which are defined by
\begin{align}
\Omega :=&\frac{j\tau (R^{6/n}-1)}{R^6\varDelta_L}\,,\\
V^\pm :=& 2\Omega L_R\pm \frac{\sqrt{
\Delta \left[j^2L_R^2+\varDelta _LR^6\left(L^2+{m^2R^2}/4\right)\right]
}}{\varDelta_LR^6} \,.\label{V_pm}
\end{align}
The allowed region is $E>V^+$ or $E<V^-$ for $\varDelta_L>0$, whereas it is
${\rm min}[V^\pm]<E<{\rm max}[V^\pm]$ for $\varDelta_L<0$.

Equation~(\ref{NH_timeeq}) becomes
\begin{align}
\frac{\D T}{\D \lambda }=\frac{4R^4\varDelta_L}{\Delta
 }\left(E-2\Omega L_R\right)\,.
\end{align}
When $\Delta>0$ and $\varDelta_L>0$, 
$E>V^0:=2\Omega L_R$ follows. Thus, 
$E$ must be positive for a particle with $\Omega L_R>0$
moving forwards with respect to the time coordinate $T$. 
Inside the VLS ($\varDelta_L<0$) where $\Delta>0$, 
a particle with $E>V^0$
moves backwards with respect to the coordinate $T$. 
One also verifies that the horizon $\Delta=0$ is an infinite redshift
surface for the time coordinate $T$, which is of course a coordinate artifact.

From~(\ref{Lsq}) one finds $L^2\ge L_R^2$. When the equality holds, 
$L_L=0$ is satisfied. Thus Eq.~(\ref{Lsq2}) implies that the 
particle motion is confined on the equatorial plane $\theta =\pi/2$ and
Eq.~(\ref{NH_phieq}) implies $\phi={\rm constant}$. 
The same remark applies to the original metric~(\ref{sol}) since 
this assertion only comes from the ${\rm U}(2)$-symmetries of the
solution. 

It is clear from Eq.~(\ref{NH_radeq2}) that massless particles with $L=L_R=0$
cannot cross the VLS.  
In the over-rotating case, the geodesics with $L_R=0$ cannot cross the VLS
either, since the right-hand side of (\ref{NH_radeq2}) 
becomes negative before the VLS is reached.

In the case of $L_R \ne 0$, it is dependent on the parameters whether
the geodesic particle moving forwards can cross the VLS or not.
When $j<1 $, 
$\Omega L_R$ diverges positively (negatively) as $R\to R_L+0$ for the
particle having the  opposite (same) spin as the black hole. Hence the
particle with opposite angular momentum ($jL_R<0$) cannot penetrate the VLS
for $j<1$. 
Similarly, when $j>1$   
$\Omega L_R$ diverges  positively (negatively) as $R\to R_L+0$ for the
particle having the same (opposite) spin as the black hole. Thus the
particle with $j>1$ never penetrate the VLS when it has the same spin as
the hole $jL_R>0$.

Though causal geodesics may cross the VLS, it is shown that 
they never encounter the singularity at $R=0$ at least for $n=2, 3$. 
For $L>|L_R|$, the function inside the square-root of $V^\pm$~(\ref{V_pm})
becomes negative around $R=0$, so that $V^\pm$ does not exist around
$R=0$ and has a confluent point inside the VLS, which  
prohibits geodesics to enter inside.   
For $L=|L_R|$, it can be easily shown that 
$V^+<V^0<V^-$ holds around $R=0$ and they take value $2\tau L_R/j$ at
$R=0$. It follows that geodesics with $E= 2\tau L_R/j$ may reach $R=0$. 
For $n=1$ this is indeed the case. By contrast, for $n=2, 3$ 
$\D V^0/\D R <(>)0$ holds around $R=0$ for $jL_R>(<)0$, 
which forbids the geodesics to hit the singularity since $E<V^0$ and
$V^+<E<V^-$ are the allowed region for the future-pointing particles.
Accordingly, the singularity $R=0$ has a repulsive nature.
We can expect that geodesics rarely reach the singularity 
also in the dynamical settings.

\subsection{Global structure}

We are now ready to discuss the global structures of the
time-dependent and rotating  spacetime~(\ref{sol}). 
The most useful visualization of the causal structure of a spacetime is
the conformal diagram. To this end it is necessary to find a
two-dimensional (totally geodesic) integrable submanifold.  
Now the spacetime is regarded as an $\mathbb R^2$ bundle over $S^3$. 
Unfortunately, the distribution spanned by
$\partial/\partial t$ and $\partial/\partial r$ is not integrable, 
forbidding us to have a foliation by a two-dimensional conformal diagram. 
The frame-dragging effect inevitably drives the $\psi$-motion.

Nevertheless, the two-dimensional metric
\begin{align}
\D s_2 ^2 =-\frac{\tau^2 f^2}{\Delta _L}\D t^2 +\frac{\D r^2 }{f} \,,
\label{2dmetric}
\end{align}
still contains some information about the spacetime structure and
gives us useful visualization~\footnote{For $n=3$, the 2-dimensional 
metric~(\ref{2dmetric}) is locally isometric to the static metric 
$\D s_2^2=-(\Delta/\Delta_L)\D T'^2 +\Delta ^{-1}\D R^2$, 
where $R=(tr^2+1)^{1/2}$ and $\D T' =\D t/t+R^3\Delta_L\D R/[2\Delta (R^2-1)]$.
}.
The above metric~(\ref{2dmetric}) is associated with the null geodesics
with $\theta =\pi/2$ and $\phi={\rm constant}$ corresponding to
$L=L_R=L_L=0$: hence they cannot penetrate the VLS, which is found to be
a timelike or null surface. 
As in the BMPV case, the 2-dimensional metric~(\ref{2dmetric}) is not
Lorentzian inside the VLS.

From the analysis of the previous subsection, 
we found that the original time-independent metric~(\ref{NHmetric2}) admits 
Killing horizons at  $\Delta=0$. In the non-rotating case ($j=0$),  
the null surfaces $r=0$ with $t=\pm \infty $ are
Killing horizons also for the original spacetime~\cite{MNII}, 
since the Killing vector is parallel to the generators of horizons.

When a rotation is present we must be careful. 
Now there exists a VLS~(\ref{VLS}), which is bounded below 
when $j>1$ (see left and middle plots in figure~\ref{fig:VLS}), so that 
the past horizon $t\to -\infty $ may not exist (since we are focusing on
the 2-dimensional metric~(\ref{2dmetric}), no causal geodesics penetrate
the VLS: inside the VLS is not the physical region of spacetime).  
Even if the near-horizon metric~\cite{MNII} admits some Killing
horizons, we cannot immediately conclude that they are also Killing
horizons in the original metric.

The analysis of singularities, asymptotic infinity, 
behaviors of VLS (figure~\ref{fig:VLS}) and the near-horizon geometries
have provided us  sufficient information to deduce Carter-Penrose diagrams.  
As a striking confirmation we have solved the geodesic equations numerically and obtained the 
conformal diagrams displayed in figure~\ref{fig:PD}, which may be
summarized as follows (we have excluded the special case of the
degenerate horizons).

\noindent
(i) $n=1$. 
The asymptotic region is approximated by an FLRW universe obeying a
decelerating expansion 
$a =(\bar t/\bar t_0)^{1/4}$ caused by a massless scalar field.
Then the null infinity $\mas I^-$ 
possesses an ingoing null structure. When $j<j_{(1)}$, 
two Killing horizons $R_\pm$ arise~(\ref{Rpm_1}). Since 
the VLS diverges negatively as $r\to 0$ when $j<1$  (right plots in
figure~\ref{fig:VLS}),  the conformal diagram is (I). 
Even if two horizons exist in the near-horizon geometry for $1<j<j_{(1)}$, the VLS  
conceals the past horizon $R_-$ (corresponding to $t\to-\infty $) 
since the VLS diverges positively as $r\to $ (left plots in
figure~\ref{fig:PD}).  
Then diagram (II) is obtained. 
Note that the $R_-={\rm constant}$ surface
asymptotically approaches null as $t\to\infty $, and $R_{L}$ is
timelike almost everywhere (it happens to be null precisely at one point). 
For the over-rotation $j>j_{(1)}$, no Killing horizons arise. 
Hence the conformal diagram is (V).

\bigskip
\noindent
(ii) $n=2$. 
The spacetime approaches to the marginally accelerating universe,
expanding linearly with cosmic time
$a =\bar t/\bar t_0$. This is caused by the fluid with equations of
state $P=-\rho /2$. For $\tau >1/2$ there exists two Killing
horizons~(\ref{Rpm_1}), so that conformal diagram is the same as case
(i); it is (I) for $0<j<1$, (II) for $1<j<j_{(2)}$ and (V) for $j>j_{(2)}$.   
An essential difference from the $n=1$ case arises when $\tau \le 1/2$,
in which case there exists an internal null infinity $\mas I^+_{\rm in}$ where
$R\to \infty $ with $r\to 0$ and
$t\to \infty $.  Only ingoing null particles can get to 
$\mas I^+_{\rm in}$. 
The existence of internal null infinity can be shown by solving the
geodesics asymptotically as in 4-dimensions~\cite{MNII}. 
It follows that conformal diagrams for  $\tau \le 1/2$ are 
(III) when $j<1$ and (IV) when $j>1$.

\bigskip\noindent
(iii) $n=3$. 
The conformal diagrams are similar to the Kerr-de Sitter spacetime. 
Infinity $\mas I^+$ consists of a spacelike slice due to the 
acceleration of the universe. First,  consider the case in which 
the near-horizon metric~(\ref{NHmetric2}) admits three distinct horizons 
$R_\pm$ and $R_c$. This occurs when $\sqrt{3/2}<\tau <3\sqrt 3/4$ with 
$j_{(3)-}<j<j_{(3)+}$ and $\tau>3\sqrt{3}/4$ with $(0\le)j<j_{(3)+}$. 
Taking into account the fact that 
for $j<1$ the VLS $t_{\rm VLS}(r)$ diverges negatively as $r\to 0$, which
removes past horizons ($t\to-\infty $ and $r\to 0$ with $t r^2$ finite)
in the near-horizon geometry~(\ref{NHmetric}).  
Therefore when $\tau>3\sqrt{3}/4$ with $(0\le)j<1$
the conformal diagram is (VI), whereas it is
(VI') when $1<j<j_{(3)+}$ with $\tau>3\sqrt 3/4$, or 
$(1<)j_{(3)-}<j<j_{(3)+}$ with $\sqrt{3/2}<\tau <3\sqrt 3/4$. 
These two are essentially the same: they constitute the different
coordinate  patches depending on the value of $j$. 
In (V') the slice $t=0$ and $r\to \infty $ with $tr^2 $
 finite comprises a null boundary.  
When there appears only a cosmological horizon $R_c$ (i.e., 
$\tau <\sqrt{3/2}$, $\sqrt{3/2}<\tau <3\sqrt 3/4$ with 
$j<j_{(3)-}$ or $j>j_{(3)+}$ and $\tau>3\sqrt{3}/4$ with $j>j_{(3)+}$), 
the spacetime diagram is (VII) for $j<1$ and (VII') otherwise.  
Again, (VII) and (VII') are essentially identical. 
In (VII') the slice $t=0$ and  $r\to \infty $ with $tr^2$
 finite is also a null surface.

To summarize, the cases (I), (II), (VI) and (VI') correspond to the
rotating black-hole geometry.

\begin{widetext}
\begin{center}
\begin{figure}[h]
\includegraphics[width=15cm]{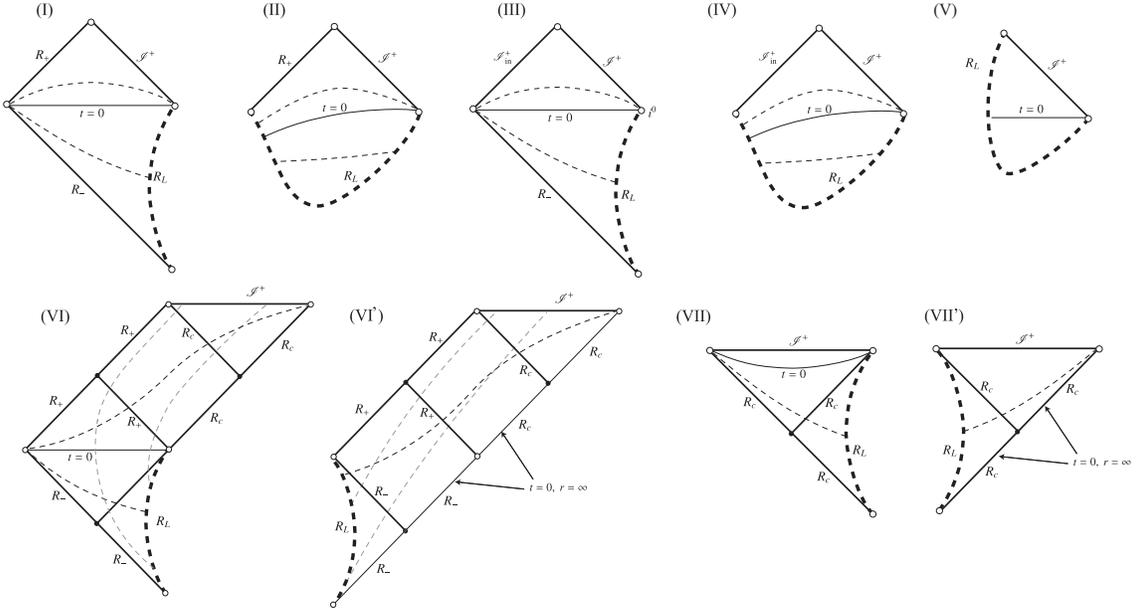}
\caption{Conformal diagrams of the 2-dimensional spacetime 
(\ref{2dmetric}), by which null geodesics with zero angular momentum 
is described.
$R_+$, $R_-$ and $R_c$ are all Killing horizons corresponding
 respectively to the black-hole event horizon, the white-hole horizon and
 the cosmological horizon. The thick dotted curves represent the VLS. 
Thin black and gray dotted curves are $t={\rm constant}$ and 
$r={\rm constant}$ surfaces, respectively.   
White and black circle are infinities (including throat) and bifurcation 
surfaces.
Since the 2-dimensional  metric~(\ref{2dmetric}) becomes Riemannian inside 
the VLS, 
the diagrams come to an end at $R_L$. Remark that we are formally
 writing the 2-dimensional figures, there still remains the angular
 motion because of the frame dragging: these figures do not display all
 the causal information. 
Though these diagrams are restricted to the $r^2>0$ region, 
the spacetime can be extended across the null surfaces $R_\pm$
and $R_c$, which are nothing but the ordinary chart boundaries.
 The conformal diagrams are 
(I) for $n=1$ with $j<1$, and for $n=2$ with $j<1$ and $\tau >1/2$, 
(II) for $n=1$ with $1<j<j_{(1)}$, and for $n=2$ with $1<j<j_{(2)}$ and
 $\tau >1/2$,  
(III) for $n=2$ with $\tau \le 1/2$ and $j<1$,
(IV) for $n=2$ with $\tau \le 1/2$ and $j>1$, 
(V) for $n=1$ with $j>j_{(1)}$ and for $n=2$ with 
$\tau >1/2$ and $j>j_{(2)}$, whereas diagrams
(VI)--(VII') correspond to $n=3$: 
(VI) for $\tau >3\sqrt 3/4$ with $j<1$, 
(VI') 
for $\tau >3\sqrt 3/4$ with $1<j<j_{(3)+}$, and 
for $\sqrt{3/2}<\tau<3\sqrt 3/4$ with $j_{(3)-}<j<j_{(3)+}$, 
(VII) for $\tau <\sqrt{3/2}$ with $j<1$ and
for $\sqrt{3/2}<\tau<3\sqrt 3/4$ with $j<1$, 
and
(VII') for $\tau <\sqrt{3/2}$ with $j>1$, 
for $\sqrt{3/2}<\tau<3\sqrt 3/4$ with $1<j<j_{(3)-}$ or 
$j>j_{(3)+}$ and 
for $\tau>3\sqrt 3/4$ with $j>j_{(3)+}$.
}
\label{fig:PD}
\end{figure}
\end{center}
\end{widetext}

\section{Dimensional oxidization and reduction}
\label{sec:liftup}

In the previous sections, some black hole solutions in the STU theory 
have been elaborated in the framework of the 5-dimensional theory. 
We shall discuss in this section the liftup and compactification
procedure to other number of dimensions.

\subsection{Lift up to M-theory}

The time-evolving and spatially-inhomogeneous solutions in 4-
and 5- dimensions were originally derived from the dimensional reduction
of intersecting M-branes in 11-dimensional supergravity. Now we argue
the solutions of case (ii)--where
two of $V_I$'s vanish--can be embedded in 11-dimensional supergravity.

The 11-dimensional supergravity action  is given by
\begin{align}
S_{11}=\frac{1}{2\kappa_{11}^2} \int\left(
{}^{11}R\star_{11} 1 -\frac 12 {\ma F}\wedge
 \star_{11}{\ma F}-\frac 16 \ma A \wedge \ma F\wedge \ma F \right)\,,
\label{11Daction}
\end{align}
where $\ma F=\D \ma A$ is the 4-form field strength.
The equations of motion are Einstein's equations, 
\begin{align}
{}^{11}R_{AB }-\frac 12 {}^{11}R g_{AB}=&\frac{1}{2\cdot 3!}\left(
\ma F_{ACDE}{\ma F_B }^{CDE}
\right. \nonumber \\& \left.
-\frac 18
g_{AB }\ma F_{CDEF}\ma F^{CDEF
 }\right)\,,
\label{11DEOM}
\end{align}
and the gauge-field equations
\begin{align}
 \D \star_{11}\ma F +\frac 12 \ma F\we \ma F&=0\,.
\label{11DMaxwell}
\end{align}
 In this section $A,B,...$ denote the 11-dimensional  indices.

Let us consider the ``intersecting M2/M2/M2 metric'' of the following
form~\cite{Elvang:2004ds},
\begin{align}
\D s_{11}^2= & \D s_5^2+X^1 \left(\D y_1^2+\D y_2^2 \right)
+X^2 \left(\D y_3^2+\D y_4^2 \right) 
\nonumber \\
& +X^3 \left(\D y_5^2+\D y_6^2 
\right)\,,\label{11Dmetric}\\
\ma A=& A^{(1)} \we  \D y_1 \wedge \D y_2 +
A^{(2)}\we \D y_3\wedge \D y_4 \nonumber \\
&+A^{(3)}\we \D y_5\wedge \D y_6\,,
\end{align}
where the metric is independent of the brane coordinates
$y_1, ...,y_6$. 
This solution is specified by 5-dimensional metric, 
\begin{align}
\D s_5^2 = &-\left(H_1H_2H_3\right)^{-2/3} \left(\D t+\omega \right)^2
\nonumber \\
& +\left(H_1H_2H_3\right)^{1/3}h_{mn }\D x^m\D x^n \,,
\label{11Dmetric2}
\end{align}
as well as 
three scalars $X^I~(I=1,2,3)$ and three one-forms $A^{(I)}$ which are
given by
\begin{align}
A^{(I)} =& H_I^{-1} \left(\D t+\omega \right)\,,\qquad 
X^I =H_I^{-1}\left(H_1H_2H_3\right)^{1/3}\,.
\end{align}
Here, $h_{mn}$ is the metric on the 4-dimensional base space. 
$\omega=\omega_m \D x^m$ is viewed as a one-form on the base space, 
i.e., $\omega_\mu V^\mu =0$ where $V^\mu =(\partial /\partial t)^\mu $.

Since the metric ansatz~(\ref{11Dmetric2}) is independent of the
coordinates $y_1, ..., y_6$, the solution can be dimensionally reduced
to 5-dimensions. 
Noting that the six-torus ${T}^6$ has a constant volume
$X^1X^2X^3=1$, it turns out that the 5-dimensional metric $\D s_5^2 $  
is the 5-dimensional Einstein-frame metric.  
Thus, the metric ansatz~(\ref{11Dmetric}) gives 
the 5-dimensional  action of gravity sector as
\begin{align}
S_g &=\frac{1}{2\kappa_5^2 }\int \D ^{5}x\sqrt{-g_5} 
\left[{}^5 R -\frac 12 \sum_I(\nabla^\mu \ln X^I)
(\nabla_\mu \ln X^I)\right]\,,
\label{5Deffection}
\end{align}
where we have used $X^1X^2X^3=1$.
We can proceed the form field sector analogously. Letting 
$F^{(I)}:=\D A^{(I)}$ denote the two-form field strengths, we find
\begin{align}
&\ma F_{ABCD }\ma F^{ABCD}  
= 12\sum_I (X^I)^{-2 } F^{(I)}_{\mu \nu }F^{(I)\mu \nu }\,,\\
&\ma A\wedge \ma F \wedge \ma F =2\left(A^{(1)}\we F^{(2)}\we F^{(3)} 
+A^{(2)}\we F^{(3)} \we F^{(1)}
\right.\nonumber \\ &\qquad \qquad \qquad  \left.
+A^{(3)}\we F^{(1)} \we F^{(3)} \right)\wedge {\rm Vol}(T^6)\,,
\end{align}
then  the Lagrangian for the gauge fields reads 
\begin{widetext}
\begin{align}
S_F =\frac{1}{2\kappa_5^2 } \int \D ^5 x \sqrt{-g_5}
\left[
-\frac{1}{4} \sum_I (X^I)^{-2 } F^{(I)}_{\mu \nu }F^{(I)\mu \nu }
+\frac{1}{12}\epsilon^{\mu\nu\rho\sigma\tau}
\left(A^{(1)}_{\mu }F^{(2)}_{\nu\rho }F^{(3)}_{\sigma\tau}+
A^{(2)}_{\mu }F^{(3)}_{\nu\rho }F^{(1)}_{\sigma\tau  }+
A^{(3)}_{\mu }F^{(1)}_{\nu\rho }F^{(2)}_{\sigma\tau  }\right)
\right]\,,
\end{align}
\end{widetext}
where 
$\epsilon_{\mu\nu\rho\sigma\tau}$  
is the volume-element 
compatible with the 5-dimensional metric $\D s^2_5$ and 
$\kappa_5^2:=\kappa_{11}^2/{\rm Vol}(T^6)$. 
It follows  that the reduced action $S_5=S_g+S_F$ 
exactly coincides with that of the STU-theory; the 
5-dimensional minimal ungauged ($\mathfrak g=0$) ${\rm U}(1)^3$-supergravity~(\ref{5Daction})
with the metric of the potential space given by
\begin{align}
 G_{IJ}=\frac 12 {\rm diag}\left[(X^1)^{-2}, (X^2)^{-2}, (X^3) ^{-2} \right]\,,
\end{align}
and the constants $C_{IJK}$ are totally-symmetric in ($IJK$)
with $C_{123}=1$ and $0$ otherwise.

If we consider three equal harmonics $H_1=H_2=H_3:=H$
(i.e., $X^I=1$, $A^1=A^2=A^3=:(2/\sqrt 3)A$ and $F=\D A$), 
all scalar fields are trivial. Then 
the action $S_5=S_g+S_F$ reduces to 
that of the minimal supergravity in 5-dimensions~\cite{GGHPR},
the action of which is given by
\begin{align}
S_5=\frac{1}{2\kappa_5^2}
&\int \D ^5x \sqrt{-g_5}\left({}^5 R -F_{\mu \nu }F^{\mu \nu }
\right. \nonumber\\&\left.
+\frac{2}{3\sqrt 3}\epsilon^{\mu\nu\rho\sigma\tau }A_\mu F_{\nu \rho }
F_{\sigma\tau }\right)\,.
\label{5DminimalSUGRA}
\end{align}

\subsubsection{Supersymmetric solution in ungauged theory}

Let us first consider the case where the 5-dimensional spacetime is supersymmetric, 
i.e., there exists a nontrivial Killing spinor satisfying
(\ref{KS_STU1}) and (\ref{KS_STU2}) with $\mathfrak g=0$~\cite{GR,Elvang:2004ds}. 
For the timelike family of solutions for which $V=\partial/\partial t$
is a timelike Killing  vector,
the supersymmetry requires that the base space is hyper-K\"ahler and
the Maxwell fields are expressed as 
\begin{align}
 F^{(I)}=\D [fX^I(\D t+\omega )]+\Theta ^{I}\,,
\end{align}
where $\Theta^{I}$ are self-dual 2-forms on the base space satisfying 
$X_I\Theta ^{I}=-f(\D \omega+\star_h\D \omega )/3$.  The Bianchi
identity for $F^{(I)}$ requires $\D \Theta^{I}=0$, and the Maxwell
equation leads to 
\begin{align}
 {}^h\Delta (f^{-1}X_I) =\frac 1{12} C_{IJK}\Theta ^{(J)mn}\Theta^{(K)}_{mn}\,. 
\end{align}
For $\Theta^I=0$, the solution reduces precisely to the one 
assumed for the pseudo-supersymmetric solutions (\ref{f_cube}).

If we set $h_{mn}=\delta _{mn}$, $\omega =0$ and $H_I=1+Q_I/r^2$, 
the metric describes the standard static intersecting M2/M2/M2-branes
with corresponding charges $Q_I$. 
In this case 
the 11-dimensional solution  admits a Killing spinor
$\varepsilon =(H_1H_2H_3)^{-1/6}\varepsilon_\infty $ with 
\begin{align}
 i\Gamma ^{0\hat y_1\hat y_2}\varepsilon_\infty  =\varepsilon_\infty \,,~
 i\Gamma ^{0\hat y_3\hat y_4}\varepsilon_\infty =\varepsilon_\infty \,,~
 i\Gamma ^{0\hat y_5\hat y_6}\varepsilon_\infty =\varepsilon_\infty \,.
\end{align}
satisfying 
\begin{align}
 \left[
{}^{11} \ma D_A  +\frac{i}{288}\left({\Gamma_A }^{BCDE}-8
{\delta_A }^B \Gamma^{CDE }\right) \ma F_{BCDE
 }\right]\varepsilon =0\,,
\end{align} 
where $\Gamma $ is the 11-dimensional gamma matrices. 
It deserves to mention that 
the fact that $\D s_5^2$ in~(\ref{11Dmetric}) is the 5-dimensional
Einstein-frame metric means that the causal pathologies are not cured
by lifting up to M-theory.

\subsubsection{Dynamically intersecting M2/M2/M2-branes}
\label{M2M2M2}

Let us next consider the non-supersymmetric case 
where the metric is time-dependent. 
The importance of dynamically intersecting branes in supergravity theory 
lies in their applications to cosmology and dynamical black holes.  
The dynamically intersecting branes without rotation are analyzed in
detail in~\cite{MOU}. We are going to discuss its rotating version.

The potential 
$V=27C^{IJK}V_IV_JX_K$ vanishes identically for the STU theory with 
the case (ii) $V_1\ne 0, V_2=V_3=0$.  
This lies at the heart of why the pseudo-supersymmetric solution
of the case (ii)  derived in section~\ref{5D_sol_SG} [see Eq. (\ref{sol_ii})]
can be embedded in to 11-dimensional supergravity. 
Note, however, that the 11-dimensional
configuration is no-longer (true nor fake) supersymmetric. 
Nevertheless, the 5-dimensional pseudo-supersymmetry 
justifies the mechanical equilibrium of dynamically intersecting branes. 
If $\bar H_1, H_2$ and $H_3$ represent harmonics with a single point
source on the Euclid 4-space,  
the solution describes the dynamically intersecting rotating M2/M2/M2
branes obeying the harmonic superposition rule. 
For the vanishing charges $\bar H_1=0$ and $H_2=H_3=1$, 
the background metric is obtained, which is 
the 11-dimensional ``rotating'' Kasner universe, 
\begin{widetext}
\begin{align}
\D s_{11}^2=&-\left[\D \bar t+\frac{J}{2r^2(\bar t/\bar
 t_0)^{1/2}}(\sin^2 \vartheta \D \phi_1+\cos^2\vartheta \D \phi_2)\right]^2+
(\bar t/\bar t_0)^{1/2}\left[\D r^2+r^2(\D \vartheta^2+\sin^2\vartheta \D \phi_1^2+
\cos^2\vartheta \D \phi_2^2)\right] \nonumber \\
&+
(\bar t/\bar t_0)^{-1}\left(\D y_1^2+\D y_2^2\right)+(\bar t/\bar t_0)^{1/2}
\left(\D y_3^2+\cdots +\D y_6^2\right)\,.
\end{align}
\end{widetext}
Here $\bar t\propto t^{2/3}$ measures the cosmic time. The 11-dimensional universe
collapses into the $y_1$-$y_2$ directions and expand other
 directions~\cite{Gibbons:2005rt}.
It follows that the three
kinds of branes are intersecting in the background of Kasner universe. 
The case of $J=0$ recovers the conventional vacuum Kasner solution.

\subsubsection{The cases {\rm (iii)} and {\rm (iv)}}

For the cases (iii) and (iv), there exists a non-zero potential 
in the fake supergravity theory. 
It might be reasonable to expect that the FLRW universe may be realized 
from the viewpoint of intersecting branes, which are the fundamental constituents of
supergravity.  
Assuming the brane intersection rule~\cite{MOU} and 
making the nonzero vacuum expectation values of the 4-form $\ma F$, 
we have tried to uplift the 
solutions~(\ref{sol}) with (\ref{sol_H}) into 11-dimensions but failed. 
Whether the present solutions are obtainable from the brane
picture is an outstanding issue at present. 
We leave this possibility to the future work.

\subsection{Compactification to 4-dimensions}

When discussing the FLRW spacetime, it is much more reasonable to 
argue within the 4-dimensional effective theory.
In this section we shall show how to achieve this.

\subsubsection{Dimensional reduction via Gibbons-Hawking space
 and Kaluza-Klein black hole}

One can obtain the 4-dimensional solutions in~\cite{GMII,MNII} via
dimensional reduction of 5-dimensional solutions~(\ref{SUSYmetric}) as follows.  
We employ the Gibbons-Hawking
space~\cite{Gibbons:1979zt} as a 4-dimensional base space, 
\begin{align}
 \D s^2_{\mathcal B} = {h}^{-1}\left(\D x^5 +\chi_i\D x^i \right)^2 +h 
\delta_{ij }\D x^i \D x^j\,, 
\label{GHmetric}
\end{align}
where $i,j,...$ denote 3-dimensional indices (hence no distinction is
made for upper and lower indices) and 
\begin{align}
 \vec \nabla \times \vec \chi =\vec \nabla h\,.\label{chi}
\end{align}
$\vec \nabla$ is the derivative operator on the flat 
 Euclid 3-space and usual vector convention will be
used for the quantities on the Euclid space henceforth.  The integrability condition of (\ref{chi}) 
implies that $h$  is a harmonic function on the Euclid space 
$\vec \nabla^2 h=0$.   
In the Gibbons-Hawking base space, 
$\partial/\partial x^5$ is a Killing vector preserving the 
three complex structures, which are given by~\cite{Gibbons:1987sp}
\begin{align}
\mathfrak J^{(i)} =(\D x^5+\chi )\we \D x^i-\frac 12 h\epsilon_{ijk}\D
 x^j\we \D x^k \,.
\end{align}
The orientation is chosen in such a way that the complex structures are
anti-self-dual, viz, the volume form is given by 
$h\D x^5 \wedge \D x^1 \wedge \D x^2\we \D x^3$. 
Under the change of Killing coordinates $x^5 \to x^5 +g(x^i)$ where $g$
is an arbitrary function of $x^i$, $\chi_i$ transforms as 
$\chi_i \to \chi_i-\partial_i g$ and $h$ is unchanged 
in order to preserve the metric form.  
Prime examples of Gibbons-Hawking space are the flat space 
($h=1$ or $M/|{\vec x}|$), the Taub-NUT space ($h=1+M/|{\vec x}|$)
and the Eguchi-Hanson space ($h=M/|{\vec x-\vec x}_1|+M/|{\vec x-\vec x}_2|$).

Assuming that the vector field $\partial/\partial x^5$ is also a
Killing vector for the whole 5-dimensional spacetime, 
it turns out that functions $H_I$ are also harmonics on the Euclid 
3-space $\vec \nabla^2 H_I=0$ (and the linear time-dependence remains
intact).  Let $\omega $ decompose as
\begin{align}
\omega =\omega_5(\D x^5+\chi_i\D x^i)+\omega_i\D x^i \,, 
\end{align}
and let us write the metric as  
\begin{align}
 \D s_5^2= &\Lambda \left[\D x^5+\chi_i \D x^i-f^2
 \omega_5\Lambda^{-1} (\D t+\omega_i\D x^i )\right]^2
\nonumber \\ &
-
fh^{-1}\Lambda ^{-1}(\D t+\omega_i \D x^i)^2 +f^{-1}h\delta_{ij}\D x^i\D x^j\,,\\
=&:e^{-4\sigma /\sqrt 3}(\D x^5+B_\alpha \D x^\alpha )^2 +e^{2\sigma/\sqrt
 3}g_{\alpha\beta }\D x^\alpha \D x^\beta \,,
\label{DD}
\end{align}
where $\Lambda=f^{-1}h^{-1}-f^2\omega_5^2 $, 
$g_{\alpha\beta }$ is the 4-dimensional Einstein frame metric, 
$
B_\alpha \D x^\alpha \D x^\alpha 
=\chi_i \D x^i-f^2\omega_5\Lambda^{-1}(\D t+\omega_i\D x^i )
$ 
 is the Kaluza-Klein gauge field and
$\sigma=-{\sqrt{3}\over 4} \ln \Lambda$ is a dilaton field. 
The anti-self duality of Sagnac curvature
$\D \omega+\star_h\D \omega=0 $~\cite{GibbonsBMPV} reduces to 
\begin{align}
\vec \nabla \times \vec \omega =h^2 \vec \nabla (h^{-1}\omega_5 )\,. 
\label{3D_omega} 
\end{align}
The integrability condition of this equation is 
$\vec\nabla^2 \omega_5=0$, i.e., $\omega_5 $ is another harmonic
function.  
The Einstein frame metric $g_{\alpha \beta }$ is given by 
\begin{align}
\D s_4^2&
=-\Xi (\D t+\omega_i \D x^i)^2+ \Xi^{-1} \delta_{ij}\D x^i \D x^j\,,
\end{align}
with
\begin{align}
\Xi :=fh^{-1}\Lambda ^{-1/2}\,,
\end{align}
where $\vec\omega $ is determined by~(\ref{3D_omega}) up to a gradient.

When $\omega_5$ is proportional to $h$, Eq.~(\ref{3D_omega}) implies
that $\vec \omega $ is written as a gradient of some scalar function,
which can be made to vanish by redefinition of $t$ and harmonic
functions if we work in a ``Coulomb gauge'' 
$\vec \nabla \cdot \vec \omega =0$. Thus   
the 4-dimensional rotation vanishes ($\vec \omega=0$) in this case. If 
two harmonics are equal ($H_2=H_3$) in the STU-theory and $\omega_5=0$, 
the 4-dimensional solutions given in~\cite{MN,MNII} except the 
$n_T=4$ case are recovered. 
Since the dimensional reduction does not spoil the fraction of
supersymmetries,  it turns out that the 4-dimensional 
solutions in~\cite{MN,MNII} are also pseudo-supersymmetric 
in the context of fake supergravity.

The resulting 4-dimensional theory involves many scalar and vector
multiplets. To see this we consider the general Kaluza-Klein ansatz~(\ref{DD}). 
Defining 
\begin{align}
&H_{\alpha\beta }=2 \partial_{[\alpha }B_{\beta ]}\,, 
\qquad A^{(I)}=A_\alpha ^{'(I)}\D x^\alpha  +\theta ^{(I)}\D x^5\,,
\nonumber \\
&F^{'(I)}_{\alpha\beta }=2\partial_{[\alpha }A_{\beta ]}^{'(I)}\,,
\quad 
{}^4F_{\alpha\beta }^{(I)}=F^{'(I)}_{\alpha\beta }-2\partial_{[\alpha
 }\theta^{(I)}B_{\beta ]}\,,
\end{align}
one finds that the 5-dimensional theory~(\ref{5Daction}) leads to 
the following 4-dimensional effective Lagrangian, 
\begin{widetext}
\begin{align}
L_4 =& {}^4R-2k^2  Ve^{2\sigma/\sqrt 3}
-2g^{\alpha \beta }\partial _\alpha \sigma\partial_\beta \sigma-\frac{1}{4}e^{-2\sqrt
 3\sigma }H_{\alpha\beta }H^{\alpha\beta }
-\ma G_{AB}g^{\alpha\beta }\partial _\alpha \phi^A\partial_\beta \phi^B
\nonumber \\
&
-\frac 12 e^{-2\sigma /\sqrt 3}
G_{IJ}{}^4 F^{(I)}_{\alpha\beta }{}^4 F^{(J)\alpha\beta }
-e^{4\sigma/\sqrt 3}G_{IJ}g^{\alpha\beta }\partial _\alpha \theta ^{(I)}
\partial_\beta \theta^{(J)}
\nonumber \\
&
-\frac 18{\epsilon^{\alpha\beta\gamma\delta }}C_{IJK}\theta ^{(I)}\left(
{}^4 F^{(J)}_{\alpha\beta }{}^4F^{(K)}_{\gamma\delta }-\theta^{(J)}\cdot
 {}^4F_{\alpha\beta }^{(K)}H_{\gamma\delta }
+\frac 13 \theta^{(J)}\theta^{(K)}H_{\alpha\beta }H_{\gamma\delta}
\right)\,. 
\end{align}
\end{widetext}
Thus the 4-dimensional effective theory derived from the 
Lagrangian~(\ref{5Daction}) comprises $2N$ scalars 
$(\sigma, \phi^A, \theta^{(I)})$ and
$N+1$ gauge fields $(A'^{(I)}_\mu , B_\mu )$ in general. 
While, its supersymmetric solution is specified by $N+2$ harmonics
($H_I, h, \omega_5$).

As an obvious application let us consider the case where the 4-dimensional base space 
($\ma B, h_{mn}$)  is the Taub-NUT space. The Taub-NUT metric can be written as a
Gibbons-Hawking form~(\ref{GHmetric}) as, 
\begin{align}
\D s^2_{\rm TN}=&\left(\varepsilon +\frac{M}{\rho }\right)^{-1} M^2 (\sigma_R^3)^2 
\nonumber \\ &+
\left(\varepsilon +\frac{M}{\rho }\right) \left[\D \rho ^2 +
\rho ^2
\left\{(\sigma_R^1)^2 +(\sigma_R^2)^2 \right\}
\right]\,.
\label{TaubNUT}
\end{align} 
where $\rho:=|\vec x|$ and  $M (>0) $ corresponds to the NUT parameter. 
For later convenience, we have introduced a parameter $\varepsilon$,
which is unity for the Taub-NUT space.

A natural 5-dimensional background ($|\vec x| \to \infty $) in this case
is 
\begin{align}
\D s_{\rm GPS}^2 = -\D \bar t^2 +a(\bar t) ^2 \D s_{\rm TN}^2 \,.
\label{GPS}
\end{align}
where the scale factor $a(\bar t)$ is given by (\ref{scale_factor1}) and
(\ref{scale_factor2}). 
This is the Gross-Perry-Sorkin type monopole~\cite{Gross:1983hb} immersed in the
FLRW universe. At large distance $|\vec x|\to \infty $ 
it may be rewritten as a U(1)-fibration over the FLRW
universe $M_5 \simeq M_4\times S^1$,  
\begin{align}
\D s_{\rm GPS}^2 =\D s^2 _{\rm FLRW} +M a(\bar t)^2 \rho  (\sigma
 _R^3)^2 \,.   
\end{align}
Thus the spacetime is effectively 4-dimensional at infinity. 
Since the metric~(\ref{GPS}) admits a homothetic Killing field,
one can analyze its causal structures analytically. The conformal
diagrams are the same as the 5-dimensional FLRW universe.

Reminding the fact that the flat Euclid space is
recovered when $\varepsilon =0$ in the metric~(\ref{TaubNUT}) (note that
in this case $M$ is not the NUT charge), 
the spacetime structure as $\rho \to 0$ 
(with or without $t\to \pm \infty $) 
is identical to that for the solution~(\ref{sol}). 
Then the vicinity of horizons is indeed 5-dimensional. 
Therefore this geometry describes a Kaluza-Klein type black
hole~\cite{Ida:2007vi}.

\subsubsection{A caged black hole}

As discussed in~\cite{Myers:1986rx,Maeda:2006hd} for the supersymmetric case, 
a caged black-hole geometry is obtained by  
superimposing an infinite number of black holes 
aligned in one direction 
with an equal separation. 
Since the present time-dependent solution found in
section \ref{5D_sol_SG}  is linearized in space, we can 
construct similar configurations easily. 
Decomposing the Euclid 4-space coordinates as $x^m=(x,y,z,w)$ with the
orientation $\D x\we \D y \we \D z\we \D w$ and
putting the same point sources along $w$-axis with an equal-spacing of 
$2\pi R_5$, we obtain 
\begin{widetext}
\begin{eqnarray}
H_S&=&1
+Q_{S} \sum_{k=-\infty}^{\infty} \frac{1}{\rho^2+(w+2\pi k R_5)^2}
=1+{Q_S\over 2  R_5^2}
{\sinh \bar \rho \over \bar \rho\left(
\cosh \bar \rho-\cos \bar{w} \right)}
\label{ha1}\,,
\\
H_T&=&{t\over t_0}
+Q_{T} \sum_{k=-\infty}^{\infty} \frac{1}{\rho^2+(w+2\pi k R_5)^2}
={t\over t_0}+{Q_T\over 2  R_5^2}
{\sinh \bar \rho\over \bar \rho\left(
\cosh \bar\rho-\cos \bar{w} \right)}
\label{ha2}\,,
\\
\omega _{\phi_1}&=& {J}
\sum_{k=-\infty}^{\infty} \frac{x^2+y^2}{[\rho^2+(w+ 2\pi k R_5)^2]^2}
={J\over 4R_5^2}
{\left(\bar{x}^2+\bar{y}^2\right)\over \bar\rho^2}
\left[
{\left(\cosh \bar\rho\cos \bar{w}-1
\right)\over \left(\cosh \bar \rho-\cos \bar{w}\right)^2}
+{\sinh \bar \rho\over \bar{\rho }
\left(\cosh \bar\rho-\cos \bar{w}\right)}
\right]\,, \\
\omega _{\phi_2}&=& {J}
\sum_{k=-\infty}^{\infty}
\frac{z^2+(w+ 2\pi k R_5)^2}{[\rho^2+(w+ 2\pi k R_5)^2]^2}
\nonumber \\
&=&
{J\over 4R_5^2}
\left[
-{(\bar{x}^2+\bar{y}^2)\over \bar{\rho }^2}
{\left(\cosh \bar\rho\cos \bar w-1
\right)\over \left(\cosh \bar \rho-\cos \bar{w}\right)^2}
+{\left(\bar \rho^2+\bar z^2 \right)\over \bar{\rho }^2}
{\sinh \bar \rho
\over  \bar\rho
\left(\cosh \bar\rho-\cos \bar{w}\right)}
\right]
\,,
\end{eqnarray}
\end{widetext}
where $\rho^2\equiv x^2+y^2+z^2$, and we have introduced 
dimension free coordinates $\bar x^m=x^m/R_5$ and 
$\bar \rho =\rho/R_5$. 
To derive these expressions we have used a series expansion
\begin{align}
\sum_{k=-\infty }^\infty \frac{1}{\xi^2+(\eta +2\pi k)^2}=\frac{\sinh \xi
 }{2\xi (\cosh \xi-\cos \eta )}\,.
\end{align}

Since this solution is periodic in the $w$-direction 
by identifying $w=0$ and $2\pi R_5$, 
it can be regarded as a deformed BMPV ``black hole''
in a compactified five-dimensional spacetime ($0 \leq w \leq 2 \pi R_5$)
with pseudo-supersymmetry.

Introducing the 3-dimensional spherical coordinates
($\rho,\Theta,\Phi$), 
which are defined by 
\begin{align}
&&
x=\rho  \sin \Theta \cos \Phi, \quad
y=\rho  \sin \Theta \sin \Phi , \quad 
z=\rho \cos \Theta
\,,~~~~~~~~
\end{align}
the 4-dimensional Einstein frame metric 
in the asymptotic region ($\rho\gg \pi R_5$)
reads 
\begin{align}
\D \bar{s}_4^2=&-f^{3/2}\left(
\D \bar{t}+\bar \omega_\Phi \D \Phi
\right)^2
\nonumber \\
& +f^{-3/2}\left[\D \bar \rho^2+\bar\rho^2\left(\D \Theta^2+
\sin^2\Theta \D \Phi^2\right)
\right]\,,
\label{CBH_4D_metric}
\end{align}
where
\begin{eqnarray}
&&
f=\left(1+{1\over 2R_5^2}{Q_S\over \bar\rho}\right)^{-n/3}
\left({t\over t_0}+{1\over 2R_5^2}{Q_T\over \bar\rho}\right)^{-1+n/3}\,,
\nonumber \\
&&
\bar \omega_\Phi= { 1 \over 4 R_5^3} {J\over \bar\rho}
\sin^2\Theta
\,.
\end{eqnarray}
In the asymptotic limit $\rho\rightarrow \infty$,
the metric~(\ref{CBH_4D_metric}) describes an FLRW universe with 
the power exponent of the scale factor being  $p=1/(4-n)$.
One might therefore expect that this solution describes a caged black hole 
in the effective 4-dimensional FLRW universe.
However, we have to be careful to judge whether it is a black hole
or not. 
A two-black hole system in the Kastor-Traschen spacetime
[the case (iv) without rotation] will collide and merge to form a single
black hole in the contracting universe ($t_0<0$).
In the expanding universe, the solution describes the time reversal one.
Namely it corresponds to the two-white hole system, 
since one object disrupt into two objects, which is 
possible for a white hole but not for a black hole.
In the present case we have infinite numbers of point sources
before identification, so that we can expect a similar result.    
It therefore appears that the object in the expanding universe corresponds to 
a splitting ``white string'' into an array of white holes.  In order to clarify this rigorously, 
we have to analyze (numerically) the horizons of 
a multi-object system in the expanding universe. 
Especially one important question to be answered 
is whether black holes will collide 
in a contracting universe for any value of $n$.

\section{Concluding remarks}
\label{conclusion}

We have presented pseudo-supersymmetric solutions to 5-dimensional
``fake'' supergravity coupled to arbitrary ${\rm U}(1)$
 gauge fields and scalar fields.  The non-compact gaugings of 
R-symmetry correspond to the Wick-rotation of gauge coupling constant
($\mathfrak g \to ik$). 
Since the bosonic action is not charged with respect to R-symmetry, no
ghosts appear in this sector, i.e., all kinetic terms possess the
correct sign. The net effect of imaginary coupling 
produces a positive potential for the scalar fields. Hence the background spacetime is 
generally dynamical, contrary to the supersymmetric case.

The metric solves 1st-order Killing spinor equation,
which automatically guarantees that the Einstein equations and the
scalar field equations are satisfied
if the Maxwell equations are solved.  
The solution is specified by  time-dependent and time-independent
harmonics $H_I$ on a hyper-K\"ahler base space.  
This encodes the balances of forces of the solution: the gravitational
attraction is adjusted to cancel the electromagnetic repulsive force
(the scalar fields can contribute both sides depending on the potential).  
We specialized to the case in which a single point source on the Euclid
4-space and explored its physical properties.   
The solutions we found are the rotating generalizations of 
our previous solutions~\cite{GMII,MNII} describing a black hole in the FLRW universe.   
The present metric has four parameters: the Maxwell charge $Q$, the
angular momentum $J$, the number of time-dependent harmonics $n$ 
and the ratio of energy densities of the Maxwell field
and the scalar field at the horizon $\tau $. 
The spacetime approaches to the rotating ${\rm AdS}_2\times S^3$ for small radii,
while it asymptotes to the FLRW cosmology for large radii. 
So, the solution is a BMPV black hole immersed in the time-dependent background
cosmology. Except the asymptotic de Sitter case, one cannot introduce a 
stationary coordinate patch even in the single centered case.   
Though we have made some simplification, it turns out that 
the solution enjoys much richer physical properties than stationary ones.

The analysis of near-horizon geometry uncovers that the horizon is
described by a Killing horizon. Hence the ambient materials fail to
accrete onto the black hole irrespective of the dynamical background. 
This property may be attributed to the pseudo-supersymmetry. 
The ``BPS'' solution maintains equilibrium, forbidding the horizon to grow.

An important issue to be noted is that the event horizon is not
extremal in general.  This is due to the fact that the 
event horizon is not generated by the coordinate vector field 
in the metric~(\ref{SUSYmetric}). 
Furthermore the event horizon is rotating, i.e., 
the event horizon is generated by a linear combination of 
time and angular Killing vectors~(\ref{generator}). 
This is in sharp contrast to the supersymmetric BMPV black hole 
with vanishing angular velocity. 
The nonvanishing angular velocity of the horizon indicates that there 
exists an ergoregion lying strictly outside the horizon. 
The presence of an ergoregion implies the possibility of rotating energy
removal process via the Penrose process and the superradiant
scattering~\cite{Nozawa:2005eu}. 
We can find that this is indeed the case for $n=3$ as shown
in Appendix~\ref{sec:superradiance}. 
For other values of $n$, the energy of a particle and a wave is not
conserved, so it is not a straightforward issue to conclude whether such an energy
extraction process is actually realizable under a dynamical setting. 
This is an interesting future
work to be argued.

We have also revealed that rotating solutions generically suffer from
causal violation in the neighborhood of singularities. The pseudo-supersymmetry
cannot elude naked time machines. The reason is obvious: the (pseudo-)supersymmetry
variations~(\ref{fakeKS1}) and (\ref{fakeKS2}) are local, so that 
they make no direct mention of global structure of spacetime such as
closed timelike curves. In particular, the timelike singularity
$t=t_s(r)$ in the $r^2>0$ domain is repulsive.

The original time-dependent equilibrium solution was derived via 
compactification of M2/M2/M5/M5-branes in 11-dimensional
supergravity~\cite{MOU}. We discussed in section~\ref{M2M2M2} that 
the present metric with a single time-dependent harmonic function 
can be embedded into 11-dimensions, describing a dynamically intersecting 
M2/M2/M2-branes in a rotating Kasner universe. 
It is shown that the 4-dimensional solution~\cite{MOU} was also derived
from compactification of 5-dimensional solution on the Gibbons-Hawking
space.  Unfortunately, 
such a liftup procedure fails to act as a chronology protector. 
It is of particular interest to see whether it oxidizes to a causally well-behaved 
solution in 10-dimensional supergravity, as in~\cite{Herdeiro:2000ap}.  
It appears appealing to examine if the occurrence of closed timelike
curves corresponds to the loss of unitarity in the 
context of de Sitter/CFT correspondence.

{\it Note added.} 
During the completion of this work, we noticed the work
of~\cite{Gutowski:2010sx}, which classifies all the
pseudo-supersymmetric solution of the theory~(\ref{5Daction}). 
It is intriguing to examine if more general classes of solutions admit
black hole horizons in the expanding universe.

\acknowledgments{
This work was partially supported 
by the Grant-in-Aid for Scientific Research
Fund of the JSPS (No.22540291) and 
and by the Waseda University Grants for Special Research Projects. 
}

\appendix

\section{Dilatonic ``black hole'' in the FLRW universe}
\label{appA}

In the body of text, we considered several gauge fields in order 
to make the horizon area nonvanishing. 
To see this more concretely, let us consider the 
4-dimensional Einstein-Maxwell-dilaton gravity in which a single 
gauge field exists, 
\begin{align}
S=\frac{1}{2\kappa_4^2} \int \left(R\star_4 1 -2 \D \sigma  \we \star_4
\D \sigma -2 e^{-2\alpha \sigma  }F\we \star_4  F \right)\,, 
\end{align}   
where $\alpha $ is a coupling constant. The BPS equations are~\cite{Gibbons:1993xt}
\begin{align}
 \left(\ma D_\alpha +\frac{i}{4\sqrt{1+\alpha^2 }}e^{-\alpha\sigma }
\gamma^{ab}\gamma_\alpha F_{ab}\right) \epsilon &=0\,,\label{EM_KSeq1} \\
\left(\gamma^\alpha \partial_\alpha \sigma -\frac{i\alpha }{2\sqrt{1+\alpha^2 }}
e^{-\alpha\sigma  }\gamma^{ab}F_{ab}\right)\epsilon &=0\,. 
\label{EM_KSeq2}
\end{align} 
(Remark that the second term in the dilatino equation~(\ref{EM_KSeq2}) has a factor
 2-discrepancy with the result
in~\cite{Gibbons:1993xt}, which seems to be a typo.)
This theory admits a static and spherically symmetric black-hole
solution~\cite{GM}, whose BPS limit is given by 
\begin{align}
&
\D s^2 =-U^{-2 /(1+\alpha^2 )}\D t^2 +U^{2/(1+\alpha^2 )} \delta_{ij}\D
 x^i \D x^j\,,
\nonumber \\
& A=\frac{1}{\sqrt{1+\alpha^2 }U} \D t\,, \qquad 
\sigma  =-\frac{\alpha }{1+\alpha^2 }\ln U\,,\label{GMsol}
\end{align}
where $U=1+Q/|{\vec x}|$. 
This metric admits a Killing spinor
$\epsilon =U^{-1/[2(1+\alpha^2 )]}\epsilon_\infty $,  where 
$\epsilon_\infty$ denotes the constant spinor (corresponding to the
asymptotic value of $\epsilon $) satisfying 
$i\gamma^0\epsilon_\infty =\epsilon_\infty $. 
We find that any harmonic function $U$ on the flat 3-space 
solves the Maxwell equations, hence this metric describes 
a multiple configuration, which is a cousin of 
a Majumbdar-Papapetrou solution.

This solution can be immersed in an FLRW background by setting 
$U=t/t_0+\bar U(x)$ where $\bar U$ is any harmonic function, and  
by introducing a Liouville-type exponential potential 
\begin{align}
V=V_0e^{2\alpha \sigma},  \qquad V_0 =
\frac{2(3-\alpha^2 )}{(1+\alpha^2
 )^2 t_0^2}\,.
\end{align}
This is a generalization of the solution given in~\cite{HH}
to any values of $\alpha $.  
This  spacetime is dynamical and approaches to the flat FLRW universe 
filling with the fluid of equation of state $P=[(2\alpha^2-3)/3]\rho $. 
Unfortunately, the  metric fails to have a regular horizon in either
case. These solutions exhibit timelike singularities at $U=0$: the 
$r\to 0 $ limit fails to give a throat geometry and the well-defined
scaling limit does not exist either.  
This illustrates that only a single gauge
field cannot sustain a black hole.

Finally, we briefly comment on the  $\alpha=\sqrt 3$ case, 
in which the theory
can be oxidized to the 5-dimensional vacuum Einstein gravity 
${}^5 R_{\mu\nu }=0$
via the Kaluza-Klein lift~(\ref{DD}). 
When $U=1+Q/|\vec x|$, the 5-dimensional metric admits a covariantly
constant {\it null} Killing vector $V^\mu =(\partial/\partial t)^\mu $,
 hence the
spacetime describes a pp-wave.  
This means that the BPS solution~(\ref{GMsol}) belongs to the null
family of solution~(see equation (4.42) of \cite{GGHPR}), so
its time-dependent generalization $U=t/t_0+Q/|\vec x|$ does not give a black hole.

\section{Superradiance from the Klemm-Sabra solution}
\label{sec:superradiance}

We have found that the black holes preserving pseudo-supersymmetry
(\ref{sol_i})--(\ref{sol_iv}) are rotating and possess ergoregion. 
Hence we expect superradiance. 
For a spacetime which is asymptotically FLRW universe, however,
it is difficult to 
argue the wave propagation since the background is dynamical: the
particle energy with an asymptotic observer is not conserved.   
In order to discuss the superradiant phenomena without such an ambiguity,
we shall address the wave propagation in the background 
the Klemm-Sabra solution~(\ref{sol_iv}) [the case (iv)], in which
 case the particle energy with respect to an observer rest at the
cosmological horizon is conserved since we are able to introduce a
stationary coordinate patch~(\ref{KSsol}) [we shall restrict to the
under-rotating case and drop the primes in the coordinates (\ref{KSsol})].

Since the stationary Killing field for an observer rest at the
cosmological horizon becomes spacelike inside the ergoregion, 
the energy measured by that observer can be negative.
Hence if a wave is scattered off by the black hole, this negative energy
modes are excited and fallen into the black hole, allowing the outside
observer to have an amplified wave coming out of the horizon. 
 
For simplicity let us consider a
massless scalar field $\Psi$, which evolves 
according to 
\begin{align}
 \nabla^\mu\nabla_\mu \Psi =0 \,. \label{Psieq}
\end{align}
Assuming
\begin{align}
\Psi =e^{-i\omega t+im_1\phi_1+im_2\phi_2}r^{-1} R(r)\Theta 
(\vartheta)\,, 
\end{align}
the massless scalar field equation~(\ref{Psieq}) is separable. 
The angular equation is the spin-weighted spherical harmonics with spin
weight $s=(m_1-m_2)/2$. 
The angular function $\Theta $ satisfies  
\begin{align}
&\frac{1}{\sin\vartheta\cos\vartheta }\frac{\D }{\D \vartheta
 }\left(\sin\vartheta\cos\vartheta \frac{\D }{\D \vartheta }\Theta
 \right)\nonumber \\
&\qquad +\left[\ell (\ell+2)-\frac{m_1^2}{\sin^2\vartheta
 }-\frac{m_2^2}{\cos^2\vartheta }\right]\Theta =0\,,
\end{align}
where $\ell =0,1,2...$ Incidentally, 
$\Theta e^{im_1\phi_1+im_2\phi_2}$ is proportional to the Wigner
D-function, an irreducible representation of ${\rm SU}(2)$~\cite{Edmonds}. 
Note that the above angular equation does not involve $\omega $,  
contrary to the Kerr case.  

Define the tortoise coordinate $r_*$ by
\begin{align}
\D r_*=\frac{2t_0 }{r \Delta_{\rm KS} }\,, \qquad \Delta_{\rm KS} :=1-\frac{H
 r^2}{4t_0^2}+
\frac{J^2 }{4t_0^2{r}^4}\,,
\end{align} 
so that $r_*\to  \infty $ as $r\to r_c$ and 
$r_*\to -\infty $ as $r\to r_+$, where
$r_+$ and $r_c (>r_+)$ denote respectively the loci of event and 
cosmological horizons with $\Delta_{\rm KS} (r_+)=\Delta_{\rm KS}(r_c)=0$.  
It follows that the radial equation obeys the 
Sch\"odinger-type equation

\begin{align}
&\frac{\D ^2 }{\D r_*^2} R +\left[\left(\omega
 -\frac{(m_1+m_2)J}{4t_0^2r^2}\right)^2 
\right.\nonumber\\ & \left.
-\frac{\Delta_{\rm KS}
 }{4t_0^2}\left\{\ell
 (\ell+2)+4t_0^2\omega^2 +\frac{\D }{\D r}(r\Delta_{\rm KS} ) \right\}\right]R=0\,.
\end{align}

It turns out that the reflected wave is amplified than the incident wave
if the frequency lies in the superradiant regime 
\begin{align}
\frac{(m_1+m_2)J}{4 t_0^2r_c^2}<\omega <\frac{(m_1+m_2)J}{4 t_0^2
 r_+^2}\,. 
\end{align}  
Such a superradiant amplification is characteristic to a rotating black
hole with an ergoregion. 
This phenomenon does not occur for the supersymmetric black hole for
which the stationary Killing field is always timelike outside the
horizon.  This means that in the pseudo-supersymmetric case 
the energy measured by a local observer is not necessarily positive,  
which makes superradiance possible.
For the black holes in the cases (ii) and (iii), or even for 
those with arbitrary value of $n$, 
we expect that the similar superradiant phenomena occur,
although there exists a technical difficulty to define 
a particle state (or the positive frequency states) in
a time-dependent spacetime.


\end{document}